\documentclass{article}

\usepackage{arxiv}

\usepackage{natbib}
\usepackage[hyphens]{url}  

\usepackage{graphicx} 
\usepackage{subfigure}
\usepackage{caption} 
\usepackage{picinpar}
\usepackage[ruled,linesnumbered]{algorithm2e}

\usepackage{xr}
\usepackage{enumerate}
\usepackage{amsmath}
\usepackage{amsthm}
\usepackage{bm}
\usepackage{microtype}
\usepackage{makecell}
\usepackage[export]{adjustbox} 
\usepackage{booktabs} 
\usepackage{graphicx}

\usepackage[utf8]{inputenc} 
\usepackage{url}            
\usepackage{amsfonts}       
\usepackage{nicefrac}       
\usepackage[normalem]{ulem}
\usepackage{tabularx}
\useunder{\uline}{\ul}{}
\usepackage[table]{xcolor}
\usepackage{multirow}
\usepackage{comment}
\newtheorem{myThm}{Theorem}
\newtheorem{myLemma}{Lemma}

\usepackage{graphicx} 
\usepackage[table]{xcolor}
\usepackage{multirow}
\usepackage{caption} 
\usepackage[export]{adjustbox} 
\usepackage{graphicx}
\usepackage{subfigure}
\usepackage{picinpar}
\usepackage[ruled,linesnumbered]{algorithm2e}

\renewcommand{\cite}{\citep}
\usepackage{diagbox}
\usepackage{comment}
\usepackage[ruled,linesnumbered]{algorithm2e}
\usepackage{amsmath}
\usepackage{amssymb}
\newtheorem{myDef}{Definition}

\usepackage{booktabs}  
\usepackage{xcolor}
\usepackage{newfloat}
\usepackage{listings}

\title{Robust Multi-Agent Coordination via  Evolutionary Generation of Auxiliary Adversarial Attackers}

\author{%
  Lei Yuan\textsuperscript{\rm 1,2}, Ziqian~Zhang\textsuperscript{\rm 1}, Lihe Li\textsuperscript{\rm 1},
  Ke Xue\textsuperscript{\rm 1},
  Hao Yin\textsuperscript{\rm 1},
    Feng Chen\textsuperscript{\rm 1},\\\bf
    Cong Guan\textsuperscript{\rm 1},
    Lihe Li\textsuperscript{\rm 1},
    Chao Qian\textsuperscript{\rm 1},
     Yang Yu\textsuperscript{\rm 1,2,}\thanks{Corresponding Author}\\
  \textsuperscript{\rm 1} National Key Laboratory for Novel Software Technology, Nanjing University, Nanjing, China \\
  \textsuperscript{\rm 2} Polixir.ai\\
  \texttt{\{yuanl, zhangzq, xuek, yinh, chenf, guanc, lilh\}@lamda.nju.edu.cn, \{qianc,yuy\}@nju.edu.cn}
}

\usepackage{bibentry}

\date{}
\begin{document}

\maketitle

\begin{abstract}
Cooperative multi-agent reinforcement learning (CMARL) has shown to be promising for many real-world applications. Previous works mainly focus on improving coordination ability via solving MARL-specific challenges (e.g., non-stationarity, credit assignment, scalability), but ignore the policy perturbation issue when testing in a different environment. This issue hasn't been considered in problem formulation or efficient algorithm design. To address this issue, we firstly model the problem as a limited policy adversary Dec-POMDP (LPA-Dec-POMDP), where some coordinators from a team might accidentally and unpredictably encounter a limited number of malicious action attacks, but the regular coordinators still strive for the intended goal. Then, we propose \textbf{Ro}bust \textbf{M}ulti-\textbf{A}ge\textbf{n}t \textbf{C}oordination via \textbf{E}volutionary Generation of Auxiliary Adversarial Attackers (ROMANCE), which enables the trained policy to encounter diversified and strong auxiliary adversarial attacks during training, thus achieving high robustness under various policy perturbations. Concretely, to avoid the ego-system overfitting to a specific attacker, we maintain a set of attackers, which is optimized to guarantee the attackers high attacking quality and behavior diversity. The goal of quality is to minimize the ego-system coordination effect, and a novel diversity regularizer based on sparse action is applied to diversify the behaviors among attackers. The ego-system is then paired with a population of attackers selected from the maintained attacker set, and alternately trained against the constantly evolving attackers. Extensive experiments on multiple scenarios from SMAC indicate our ROMANCE provides comparable or better robustness and generalization ability than other baselines.   
\end{abstract}

\section{Introduction} \label{intro}
Recently, cooperative multi-agent reinforcement learning (CMARL) has attracted extensive attention \citep{cooperativesuvery, gronauer2022multi} and shows potential in numerous domains like autonomous vehicle teams \cite{peng2021learning}, multi-agent path finding \cite{greshler2021cooperative}, multi-UAV control \cite{yun2022cooperative}, and dynamic algorithm configuration~\cite{xue2022multiagent}. Existing CMARL methods mainly focus on solving specific challenges such as non-stationarity~\cite{papoudakis2019dealing}, credit assignment~\cite{wang2021towards}, and scalability~\cite{christianos2021scaling} to improve the coordination ability
in complex scenarios. Either value-based methods~\cite{vdn,qmix,qplex} or policy-gradient-based methods~\cite{coma,maddpg,mappo} have demonstrated remarkable coordination ability in a wide range of tasks (e.g., SMAC~\cite{smac} and Hanabi~\cite{mappo}). 
Despite these successes, the mainstream CMARL methods are still difficult to be applied in real world, as they mainly consider training and testing policy in a nondistinctive environment. Thus, the policy learned by those methods may suffer from a performance decrease when encountering any disagreement between training and testing~\cite{guo2022towards}. 

Training a robust policy before deployment plays a promising role for the mentioned problem and makes excellent progress in single-agent reinforcement learning (SARL)~\cite{moos2022robust,xu2022trustworthy}. 
Previous works typically employ an adversarial training paradigm to obtain a robust policy. These methods generally model the process of policy learning as a minimax problem from the perspective of game theory~\cite{yu2021robust} and optimize the policy under the worst-case situation~\cite{pinto2017robust, zhang2020robust, zhang2022robust}. Nevertheless, the multi-agent problem is much more complex~\cite{zhang2021multi}, as multiple agents are making decisions simultaneously in the environment. Also, recent works indicate that a MARL system is usually vulnerable to any attack~\cite{guo2022towards}. Some MARL works study the robustness from various aspects, including the uncertainty in local observation~\cite{lin2020robustness}, model function~\cite{zhang2020robust}, and message sending~\cite{xue2022mis}. The mentioned methods either focus on investigating the robustness from different aspects, or apply techniques such as heuristic rules and regularizers used in SARL to train a robust coordination policy. However, how unpredictable malicious action attacks cause policy perturbation has not been fully explored in CMARL.

In this work, we aim to develop a robust CMARL framework when malicious action attacks on some coordinators from a team exist. 
Concretely, we model the problem  as a limited policy adversary Dec-POMDP (LPA-Dec-POMDP), where some coordinators may suffer from malicious action attacks, while the regular coordinators should still try to complete the intended goal. 

Towards developing such a robust policy, we propose ROMANCE, an adversarial training paradigm based on the evolutionary generation of auxiliary attackers. Specifically, we maintain a set of attackers with high attacking quality and behavior diversity among all generated attackers to avoid the ego-system overfitting to a specific attacker, where high attack quality requires the attacker to minimize the ego-system reward, and diversity refers to generating different behaviors among attackers. A sparse action regularizer is also introduced to promote behavior diversity for different attackers. Furthermore, to prevent the attackers from being too tricky for the ego-system to complete the intended mission, we limit the total number of attacks to a fixed value. For the training of the ego-system, we pair it with a population of attackers selected from the maintained set to complete the given mission, then iteratively select and update the attacker population under the customized quality score and diversity distance. Finally, we obtain a highly robust coordination policy under different types and degrees of action perturbations.

To evaluate the proposed methods, we conduct extensive experiments on multiple maps from SMAC~\cite{smac} and compare ROMANCE against multiple baselines. Empirical results demonstrate that our proposed adversarial training paradigm can indeed obtain attackers with high attack ability and diverse behaviors. Also, the coordination policy trained against the population can achieve high robustness and generalization effectiveness with alternative numbers of attacks during testing. 
Furthermore, visualization experiments indicate how ROMANCE improves robustness under malicious action perturbations.

\section{Related Work} 
\textbf{ Multi-agent reinforcement learning (MARL)}  has made prominent progress these years~\cite{cooperativesuvery, gronauer2022multi}. Many methods have emerged as efficient ways to promote coordination among agents, and most of them can be roughly divided into policy-based and value-based methods. MADDPG~\cite{maddpg}, COMA~\cite{coma}, DOP~\cite{wang2021dop}, and MAPPO~\cite{mappo} are typical policy gradient-based methods that explore the optimization of multi-agent policy gradient methods. MADDPG applies the CTDE (Centralized Training Decentralized Execution) paradigm to train the policies and optimizes each policy via DDPG~\cite{DBLP:journals/corr/LillicrapHPHETS15}. COMA also applies the centralized critic to optimize the policy but employs a counterfactual model to calculate the marginal contribution of each agent in the multi-agent system. DOP takes a forward step to apply a centralized linear mixing network to decompose the global reward in a  cooperative system and shows performance improvement for MADDPG and COMA significantly. Recently, MAPPO applies the widely proven learning efficiency of proximal policy optimization technique in single-agent reinforcement learning into MARL. Another category of MARL approaches, value-based methods, mainly focus on the factorization of the value function. VDN~\cite{vdn} aims to decompose the team value function into agent-wise ones by a simple additive factorization. Following the Individual-Global-Max (IGM) principle~\cite{qtran}, QMIX~\cite{qmix} improves the way of value function decomposition by learning a non-linear mixing network, which approximates a monotonic function value decomposition. QPLEX~\cite{qplex} takes a duplex dueling network architecture to factorize the joint value function, which achieves a full expressiveness power of IGM. \citet{wang2021towards} recently give theoretical analysis of the IGM by applying a multi-agent fitted Q-iteration algorithm.
More details and advances about MARL can be seen in reviews~\cite{ zhang2021multi,canese2021multi,zhu2022survey}.


\textbf{Adversarial training} plays a promising role for the RL robustness~\cite{moos2022robust}, which involves the perturbations occurring in different cases, such as state, reward, policy, etc. These methods then train the RL policy in an adversarial way to acquire a robust policy in the worst-case situation. Robust adversarial reinforcement learning (RARL)~\cite{pinto2017robust} picks out specific robot joints that the adversary acts on to find an equilibrium of the minimax objective using an alternative learning adversary. RARARL~\cite{pan2019risk} takes a further step by introducing risk-averse robust adversarial reinforcement learning to train a risk-averse protagonist and a risk-seeking
adversary, this approach shows substantially fewer crashes compared to agents trained without an adversary on a self-driving vehicle
controller. The mentioned methods only learn a single adversary, and this approach does not consistently yield robustness to dynamics variations under
standard parametrizations of the adversary. RAP~\cite{vinitsky2020robust} and GC~\cite{song2022robust} then learn population-based augmentation to the Robust RL
formulation. See~\cite{ilahi2021challenges,moos2022robust} for detailed reviews, and~\cite{smirnova2019distributionally, zhang2020robust, DBLP:conf/nips/0001CX0LBH20, oikarinen2021robust, xie2022robust} for some recent advances.

\textbf{Robust MARL} has attracted widespread attention recently~\cite{guo2022towards}. M3DDPG~\cite{li2019robust} learns a minimax extension of MADDPG~\cite{maddpg} and trains the MARL policy in an adversarial way, which shows potential in solving the poor local optima caused by opponents' policy altering. In order to model the uncertainty caused by the inaccurate knowledge of the model, R-MADDPG~\cite{zhang2020robustmarl} introduces the concept of robust Nash equilibrium, and treats the uncertainty as a natural agent, demonstrating high superiority when facing reward uncertainty. For the observation perturbation of CMARL, \citet{lin2020robustness} learn an adversarial observation policy to attack the system, showing that the ego-system is highly vulnerable to observational perturbations. 
RADAR~\cite{phan2021resilient} learns resilient MARL policy via adversarial value decomposition.   ~\citet{hu2022sparse} further design an action regularizer to attack the CMARL system efficiently. \citet{xue2022mis} recently consider the multi-agent adversarial communication, learning robust communication policy when 
some message senders 
are poisoned. To our knowledge, no previous work has explored CMARL under LPA-Dec-POMDP, neither in problem formulation nor efficient algorithm design. 

Furthermore, some other works focus on the robustness when coordinating with different teammates, referring to ad-hoc teamwork~\cite{DBLP:conf/aaai/StoneKKR10, DBLP:conf/iclr/GuZH022,mirsky2022survey}, or zero-shot coordination (ZSC)~\cite{DBLP:conf/icml/HuLPF20,DBLP:conf/icml/LupuCHF21,xue2022heterogeneous}. The former methods aim at creating an autonomous agent that can efficiently and robustly collaborate with previously unknown teammates on tasks to which they are all individually capable of contributing as team members.
While in the ZSC setting, a special case of ad-hoc teamwork, agents work toward a common goal and share identical rewards at each step. The introduction of adversarial attacks makes the victim an unknown teammate with regard to regular agents, while it is even more challenging because the unknown teammate might execute destructive actions. Our proposed method takes a further step toward this direction for robust CMARL.   

\section{Problem Formulation} 
This paper considers a CMARL task under the framework of Dec-POMDP \cite{oliehoek2016concise}, which is defined as a tuple $\mathcal{M} = \langle\mathcal{N,S,A},P,\Omega,O,R,\gamma \rangle$. Here $\mathcal{N} = \{1, \dots, n\}$ is the set of agents, $\mathcal{S}$ is the set of global states, $\mathcal{A}=\mathcal A^1\times...\times\mathcal A^n$ is the set of joint actions, $\Omega$ is the set of observations, and $\gamma \in [0, 1)$ represents the discounted factor. At each time step, agent $i$ receives the observation $o^i=O(s, i)$ and outputs the action $a^i\in\mathcal A^i$. The joint action $\boldsymbol{a}=(a^1,...,a^n)$ leads to the next state $s'\sim P(\cdot|s,\boldsymbol{a})$ and a global reward $R(s,\boldsymbol{a},s')$. To relieve the partial observability, we encode the history $(o_i^1, a_i^1, \dots, o_i^{t-1}, a_i^{t-1}, o_i^t)$ of agent $i$ 
until timestep $t$ into  $\tau_i$, then with $\boldsymbol{\tau}=\langle \tau_1, \dots, \tau_n \rangle$, the formal objective is to find a joint policy $\boldsymbol{\pi}(\boldsymbol{\tau}, \boldsymbol{a})$ which maximizes the global value function $Q_{ tot}^{\boldsymbol{\pi}}(\boldsymbol{\tau}, \boldsymbol{a}) =\mathbb{E}_{s,\boldsymbol{a} }\left[\sum_{t=0}^\infty\gamma^tR(s, \boldsymbol{a})\mid s_0=s, \boldsymbol{a_0}=\boldsymbol{a}, \boldsymbol{\pi}\right]$.

We aim to optimize a policy when some coordinators from a team suffer from policy perturbation. The vulnerability of CMARL makes it difficult to tolerate an unlimited number of perturbations. To avoid the ego-system from being entirely destroyed, we assume a limited number of perturbations and formulate such setting as a LPA-Dec-POMDP:
\begin{myDef}[Limited Policy Adversary Dec-POMDP] Given a Dec-POMDP
$\mathcal{M} = \langle\mathcal{N,S,A},P,\Omega,O,R,\gamma \rangle$, we define a limited policy adversary Dec-POMDP (LPA-Dec-POMDP) $\hat {\mathcal{M}} = \langle\mathcal{N,S,A}, P, K, \Omega,$ $O,R,\gamma \rangle$ by introducing an adversarial attacker $\pi_{adv}:\mathcal{S}\times$ $\mathcal{A}\times \mathbb{N}\rightarrow\mathcal{A}$. The attacker perturbs the ego-agents' policy by forcing the agents to execute joint action $\boldsymbol{\hat a}\sim \pi_{adv}(\cdot| s, \boldsymbol{a}, k)$ such that $s'\sim P(\cdot|s,\boldsymbol{\hat a})$, $r=R(s,\boldsymbol{\hat a},s')$. Where $K\in\mathbb{N}$ is the number of attacks that meets $\sum_{t}\sum_{i\in \mathcal{N}}\mathbb{I}(\hat a_t^i\neq a_t^i)\leq K$,  and $k \le K$ indicates the current remaining attack number. 

\label{def-padec-pomdp}
\end{myDef}

To efficiently address the attacking problem, we introduce a class of disentangled adversarial attacker policies by decomposing a policy into two components: victim selection and policy perturbation, in Def.~\ref{def-decom-adv}. 
\begin{myDef}[Disentangled Adversarial Attacker Policy] For an adversarial attacker policy $\pi_{adv}$, if there exist a victim selection function $v:\mathcal{S}\times \mathbb{N}\rightarrow \Delta(\mathcal {\hat N})$ and a policy perturbation function $g: \mathcal {\hat N}\times\mathcal A\times \mathbb{N}\rightarrow \Delta(\mathcal A)$, such that the following two equations hold:
\begin{equation*}
\begin{aligned}
    &\pi_{adv}(\boldsymbol{\hat a}| s, \boldsymbol{a},k) = v(i| s,k)g(\boldsymbol{\hat a}| i,\boldsymbol{ a},k)\\
    &g(\boldsymbol{\hat a}| i,\boldsymbol{a},0) = g(\boldsymbol{\hat a}| null,\boldsymbol{a},k)= \mathbb{I}(\boldsymbol{\hat a}=\boldsymbol{a}),
\end{aligned}
\end{equation*}
where $\mathcal {\hat N}=\mathcal N \cup\{null\} $, $i\sim v(\cdot| s,k)$, then we say that $v$ and $g$ disentangle $\pi_{adv}$.
\label{def-decom-adv}
\end{myDef}

As for the policy perturbation function, many heuristic-based methods~\cite{pattanaik2017robust,tessler2019action,sun2021strongest} have been proposed to find adversarial perturbations for a fixed RL policy. A common attacking way is to force the victim to select the action with the minimum Q-values at some steps~\cite{pattanaik2017robust}. Thus, for policy perturbation function $g$, an effective and efficient form could be $g(\boldsymbol{\hat a}| i,\boldsymbol{a},k) = g(\hat{a}^i,\boldsymbol{a}^{-i}|i,\boldsymbol{a},k)=\mathbb{I}(\hat{a}^i=\arg\min_{a^i} Q^i(\tau^i,a^i))$, if $k>0$ and $i\neq null$. 
For efficiency, we only focus on disentangled attacker policy $\pi_{adv} = v\circ g$ with a heuristic-based policy perturbation in the rest of the paper, and, without loss of generality, we suppose $g$ performs a deterministic perturbation with $\boldsymbol{\hat a} =g(i,\boldsymbol{a}, k)$.

\section{Method} \label{metho} 
In this section, we will explain the design of ROMANCE, a novel framework to learn a robust coordination policy under the LPA-Dec-POMDP. We first discuss the optimization object of each attacker and then show how to maintain a set of high-quality solutions with diverse behaviors by specially designed update and selection mechanisms. Finally, we propose ROMANCE, an alternating training paradigm, to improve the robustness of CMARL agents (ego-system) under policy perturbations.

\subsection{Attacker Optimization Objective}\label{subsection1}
In this section, we discuss how to train a population of adversarial attacker policies $P_{adv}=\{\pi_{adv}^j\}_{j=1}^{n_p}=\{v^j\circ g\}_{j=1}^{n_p}$ under a fixed joint ego-system policy $\boldsymbol{\pi}$, where $n_p$ is the population size. We anticipate the attacker population under the goal of high quality and diversity, where the quality objective requires it to minimize the ego-system's return and diversity encourages attackers to behave differently.
To achieve the mentioned goal, we first show that an individual optimal adversarial attacker could be solved under a specific MDP setting and then discuss how to solve it.
\begin{myThm} 
    Given an LPA-Dec-POMDP $\hat {\mathcal{M}} = \langle\mathcal{N,S,A},$ $ P, K, \Omega,O, R,\gamma \rangle$, a fixed joint policy $\boldsymbol{\pi}$ of the ego-system and a heuristic-based policy perturbation function $g$, there exists an MDP $\bar{\mathcal{M}}=(\mathcal{\bar{S},\bar{A}},\bar{P}, \bar R, \gamma)$ such that the optimal adversarial attacker $\pi_{adv}^*$ for $\mathcal{\hat M}$ is disentangled by an optimal policy $v^*$ of $\bar{\mathcal{M}}$ and $g$, where $\mathcal{\bar S}=\mathcal{S}\times \mathbb{N}$, $\bar s=(s, k), \bar s'=(s', k'), k, k' \le K$ indicates the remaining attack budget, $\mathcal{\bar A}=\mathcal N\cup \{null\}$, $\bar R(\bar s,\bar a,\bar s')=-R(s, \boldsymbol{\hat{a}},s')$,
    \begin{equation*}
        \begin{aligned}
            \bar P(\bar s'|\bar s, \bar a)=&\begin{cases}
         0\qquad\qquad\qquad\qquad\,\, k-k'\notin \{0,1\}\\
        P(s'|s,\boldsymbol{\hat a})\mathbb{I}(\boldsymbol{\hat a}=\boldsymbol{a})\quad k-k'=0\\
        P(s'|s,\boldsymbol{\hat a})\mathbb{I}(\boldsymbol{\hat a}\neq \boldsymbol{a})\quad k-k'=1
            \end{cases} ,
        \end{aligned}
    \end{equation*}
    \label{thm_opt}
   where $\bar d(\bar s_0)=d(s_0)\mathbb{I}(k_0=K)$,  $d$ and $\bar d$ are distributions over initial state in $\hat{\mathcal{M}}$ and $\bar{\mathcal{M}}$, respectively, and $\boldsymbol{a} = \boldsymbol{\pi}(s)$, $\boldsymbol{\hat a} = g(\bar a,\boldsymbol{\pi}(s),k)$ are original and forced action of ego-system, respectively.
\end{myThm}

The intuition behind Thm.~\ref{thm_opt} is that the adversarial attacker is aimed at minimizing the reward earned by the ego-system. Under a fixed heuristic-based policy perturbation function $g$, the adversarial attacker hopes to find an optimal victim policy $v$ to help decide which agent and when to enforce the attack. The proof can be found in Appendix.
The construction of $\bar{\mathcal{M}}$ makes it possible to apply off-the-shelf DRL algorithms to solve an optimal agent $v$ of $\bar{\mathcal{M}}$, thus deriving the optimal adversarial attacker of the LPA-Dec-POMDP.

Notice that limited numbers of attack, which could only be executed $K$ times, making the action sparse in both agent and time dimension. 
Such sparse action often plays a vital role in obtaining a high reward and thus would be exploited aggressively. However, the opportunities for taking sparse action are limited. If the attacker exhausts the opportunities at the very beginning, it will lead to sub-optimality. To guide the agent to take sparse action more cautiously, Sparsity Prior Regularized Q learning (SPRQ)~\cite{pang2021sparsity} constructs a regularized MDP by assigning a low probability to sparse action based on a reference distribution and solves the task by proposing a regularized objective:
\begin{equation}
    \begin{aligned}
        \max_{v} \mathbb{E}[\sum_{t=0}^T \gamma^t(\bar R_t-\lambda D_{KL}(v(\cdot| \bar s_t), p_{ref}(\cdot)))],
    \end{aligned}
\end{equation}
where $\lambda\in\mathbb{R}$,  $p_{ref}$ is the reference distribution which assigns a small probability $\delta$ to sparse action ``attack" and $1-\delta$ to ``not attack" (i.e., $p_{ref}(\cdot) = (\frac{\delta}{|\mathcal N|}, ...\frac{\delta}{|\mathcal N|}, 1-\delta)$ in this case), and $D_{KL}$ is
the Kullback-Leibler (KL) divergence. As claimed in Proposition 3.1 and 3.2 of \cite{pang2021sparsity}, the regularized optimal policy can be obtained by:
\begin{equation}
    \begin{aligned}
        v(\bar a| \bar s) = \frac{p_{ref}(\bar a)\exp{(\frac{\bar Q_v(\bar s, \bar a)}{\lambda})}}{Z(\bar s)},
    \end{aligned}
\end{equation}
where $\bar Q_v(\bar s,\bar a)$ is the regularized Q function under $v$, and $Z(\bar s)=\sum_{\bar a\in \mathcal {\bar A}} p_{ref}(\bar a)\exp{\frac{\bar Q_v(\bar s,\bar a)}{\lambda}}$ is the partition function.

Following the proposed regularized Bellman optimality operator, we parameterize Q-function with $\phi$ and the SPRQ loss function is thus defined as follows:
\begin{equation}
    \begin{aligned}
        L_{opt}({\phi}) = \mathbb{E}[(\bar Q_{\phi}(\bar s_t,\bar a_t)-y)^2],
    \end{aligned}
    \label{loss_opt}
\end{equation}
where $y=\bar r_t+\gamma\lambda\log{(\mathbb{E}_{\bar a'_{t+1}\sim p_{ref}(\cdot)}[\exp{(\frac{\bar Q_{\phi^{-}}(\bar s_{t+1},\bar a'_{t+1})}{\lambda})}])}$, and $\phi^{-}$ are the parameters of the target network.

Applying the above technique can lead to an attacker with high attack efficiency, nevertheless, only one attacker can easily overfit to a specific ego-system type, still leading to poor generalization ability over unseen situation. Inspired by the wildly proved ability of Population-Based Training (PBT)~\cite{jaderberg2019human}, we introduce a diversity regularization objective to ensure the effective behavior diversity of the whole population. 
The divergence of action distribution under same states is a reliable proxy for measuring the diversity between policies. In this way, we use Jensen-Shannon Divergence (JSD)~\cite{DBLP:conf/isit/FugledeT04} to reflect the diversity of the population $P_{adv}(\boldsymbol{\phi}) = \{\pi_{adv}^{\phi_j}\}_{j=1}^{n_p}=\{v^{\phi_j}\circ g\}_{j=1}^{n_p}$, which can be calculated as:
\begin{equation}
    \begin{aligned}
        \text{JSD}(\{v^{\phi_j}(\cdot|\bar s)\}_{j=1}^{n_p}) =\frac{1}{n_p}\sum_{j=1}^{n_p}D_{KL}(v^{\phi_j}(\cdot | \bar s),\bar v(\cdot|\bar s)), 
    \end{aligned}
\end{equation}
where $\bar v(\bar a| \bar s)=\frac{1}{n_p}v^{\phi_j}(\bar a| \bar s)$ is the average policy. Then, for the population, the regularization objective is defined as:
\begin{equation}
    \begin{aligned}
        L_{div}(\boldsymbol{\phi}) = \mathbb{E}_{\bar s\sim  S_{a}}[\text{JSD}(\{v^{\phi_j}(\cdot|\bar s)\}_{j=1}^{n_p})],
    \end{aligned}
    \label{loss_div}
\end{equation}
where $S_{a}=\bigcup_{j=1}^{n_p} S^j_{a}=\bigcup_{j=1}^{n_p} \{\bar s|k>0,\bar a\neq null, \bar a \sim v^{\phi_j}(\cdot| \bar s)\}$ is the union set of states (attack points) chosen to be attacked by adversarial attackers. 

Considering the mentioned sparse attack and behavior diversity, our full loss function can be derived:
\begin{equation}
    \begin{aligned}
        L_{adv}(\boldsymbol{\phi}) = \frac{1}{n_p} \sum_{j=1}^{n_p} L_{opt}(\phi_j)-\alpha L_{div}(\boldsymbol{\phi}),
    \end{aligned}
    \label{loss_tot}
\end{equation}
where $L_{opt}$ and $L_{div}$ are defined in Eq.~(\ref{loss_opt}) and Eq.~(\ref{loss_div}), respectively, and $\alpha$ is an adjustable hyper-parameter to control the balance between quality and behavior diversity.

\subsection{Evolutionary Generation of Attackers}
Despite the effectiveness of PBT with the objective in Eq.~(\ref{loss_tot}), 
the ego-system may overfit to some specific types of attackers in the current population and thus forget the attacking modes occurring in the early training stage.
To avoid this catastrophic result, we attempt to further improve the coverage of adversary policy space. 

Among different methods, Quality-Diversity (QD) algorithms~\cite{qd, qd-survey-optimization} can obtain a set of high-quality solutions with diverse behaviors efficiently, which have recently been used to discover diverse policies~\cite{edocs}, generate environments~\cite{qd-env-gen} and partners~\cite{xue2022heterogeneous} in RL. As a specific type of evolutionary algorithms~\cite{eabook}, QD algorithms usually maintain an archive (i.e., a set of solutions with high-performance and diverse behaviors generated so far) and simulate the natural evolution process with iterative update and selection. 

Inspired by QD algorithms, we design specialized update and selection mechanisms to generate desired auxiliary adversarial attackers. We maintain an archive $Arc_{adv}$ with the maximum size $n_a$, where each individual (i.e., attacker) is assigned its quality score and behavior. Specifically, given an ego-agent joint policy $\boldsymbol{\pi}$, the quality score of adversarial attacker $\pi_{adv}^{\phi_i}$ is defined as the attacker's discounted cumulative return:
\begin{equation}
    \begin{aligned}
        \text{Quality}(\pi_{adv}^{\phi_i}) = \mathbb{E}_{\boldsymbol{\bar \tau}} [\sum_{t} \gamma^t\bar R(\bar s,\bar a)|\boldsymbol{\pi}],
    \end{aligned}
\end{equation}
where $\boldsymbol{\bar \tau}$ is trajectory of attacker $\pi_{adv}^{\phi_i}$, and $\bar R$ is defined in Thm.~\ref{thm_opt}. To describe the behavior of $\pi_{adv}^{\phi_i}$, we calculate the distance between it and another attacker $\pi_{adv}^{\phi_j}$:
\begin{equation}
    \begin{aligned}
        \text{Dist}(\pi_{adv}^{\phi_i},\pi_{adv}^{\phi_j}) = \mathbb{E}_{\bar s\sim S_{a}^{i,j}} [\text{JSD}(v^{\phi_i}(\cdot | \bar s),v^{\phi_j}(\cdot | \bar s)) ],
    \end{aligned}
    \label{dist}
\end{equation}
where $S_{a}^{i,j}=S_{a}^i\cup S_{a}^j$ is the attack points set.
For simplicity, we use $\text{Quality}(i)$ and $\text{Dist}(i,j)$ to denote the quality score of $\pi_{adv}^{\phi_i}$ and the distance between $\pi_{adv}^{\phi_i}$ and $\pi_{adv}^{\phi_j}$, respectively, and $i,j$ are their indexes in the archive.

In each iteration, we use fitness-based selection~\cite{blickle1996comparison} according to their quality scores to select $n_p$ adversarial attackers from the archive and interact with the ego-system. 
Next, we take the optimization step described in Eq.~(\ref{loss_tot}) as an implicit mutation operator and derive new attackers. The archive is then updated one by one by adding the newly generated attackers. We avoid adding attackers with similar behaviors to keep the archive diverse~\citep{BR-Evolution, edocs}. That is, whenever we want to add a new attacker to the archive, we first choose the most similar attacker in the archive and calculate their behavior distance. If the distance exceeds a certain threshold, the new attacker will be added. Otherwise, we keep one at random. Note that if the current archive size exceeds the capacity $n_a$ after adding the new attacker, the oldest one will be deleted. The full procedure of the mentioned process is shown in Algo.~1 in Appendix, and we refer to such an iteration as a generation.

\subsection{Robustness Training Paradigm}
After obtaining a set of attackers with high quality in attacking and high diversity in behaviors, we aim to learn a robust ego-system under the existing attackers. We first investigate LPA-Dec-POMDP with a fixed adversarial attacker $\pi_{adv}$ and how ego-agent joint policy could be optimized.
\begin{myThm}
     Given an LPA-Dec-POMDP $\hat {\mathcal{M}} = \langle\mathcal{N,S,A}, $ $P, K, \Omega,O, R,\gamma \rangle$, a fixed deterministic adversarial attacker policy $\pi_{adv}$, there exists a Dec-POMDP $\mathcal{\tilde{M}} = \langle \mathcal{N,\tilde{S},A},\tilde{P},$ $\Omega,\tilde{O},\tilde{R},\gamma\rangle$, such that the optimal policy of $\mathcal{\tilde{M}}$ is the optimal policy for $\mathcal{\hat M}$ given $\pi_{adv}$, where $\mathcal{\tilde{S} }=\mathcal{S}\times\mathbb{N}$, $\tilde{s} = (s,k)$,  $\tilde{s}' = (s',k')$, $\tilde d(\tilde s_0)=d(s_0)\mathbb{I}(k_0=K)$,  $\tilde{O}(\tilde{s},i)=O(s,i)$, $\tilde R(\tilde s,\boldsymbol{a},\tilde s')=R(s,\boldsymbol{\hat{a}},s')$,
     \begin{equation*}
        \begin{aligned}
            \tilde P(\tilde {s}'| \tilde s,\boldsymbol{a})=
            \begin{cases}
            0 \qquad\qquad\quad\,\,\,\, k-k'\notin \{0,1\}\\
            P(s'|s,\boldsymbol{\hat a})\qquad otherwise
            \end{cases},
        \end{aligned}
    \end{equation*}
    where $\tilde d$ and $d$ are distributions over initial state in $\tilde M$ and $\hat{\mathcal{M}}$, respectively, and $\boldsymbol{\hat a}=\pi_{adv}(s,\boldsymbol{a},k)$ indicates the executed joint action of the ego-system. 
    \label{thm2}
\end{myThm}

\begin{myThm}
 Given $\hat {\mathcal{M}} = \langle\mathcal{N,S,A}, P, K, \Omega,O, R,\gamma \rangle$, a stochastic adversarial attacker policy $\pi_{adv}$, there exists a Dec-POMDP $\mathcal{\tilde{M}} = \langle \mathcal{N,\tilde{S},A},\tilde{P},\Omega,\tilde{O},\tilde{R},\gamma\rangle$, such that $\forall \boldsymbol{\pi}$, we have $\tilde{V}_{\boldsymbol{\pi}}(\tilde s)\leq \hat V_{\boldsymbol{\pi}\circ\pi_{adv}}(s,k)$, where $\tilde s=(s,k)$, $\hat V_{\boldsymbol{\pi}\circ\pi_{adv}}(s,k)$ denotes the state value function in the original LPA-Dec-POMDP $\mathcal{\hat M}$, for $\forall s\in \mathcal S, \forall k \in \{0,1,...,K\}$.
 \label{thm3}
\end{myThm}

The intuition behind Thm.~\ref{thm2} is that the constructed Dec-POMDP $\tilde M$ is functionally identical to the LPA-Dec-POMDP given the fixed $\pi_{adv}$. This theorem unveils that LPA-Dec-POMDP can be viewed as a particular version of Dec-POMDP whose policy should be robust under the ``under-attack" transition and reward function. The theorem is also easy to be extended to a population version where $\pi_{adv}\sim p(P_{adv})$, where $p$ is some distribution over the population of adversarial attackers $P_{adv}$. Thm.~\ref{thm3} illustrates that, under the circumstance where $\pi_{adv}$ is a stochastic policy, the value function in the new Dec-POMDP is the lower bound of the value function of the same joint policy in the original LPA-Dec-POMDP. Related proof can be found in Appendix. The theorem reveals that by optimizing the ego-system under the constructed Dec-POMDP $\mathcal{\tilde M}$, we can get a robust ego-system under the Limited Adversary Dec-POMDP $\mathcal{\hat M}$.

Accordingly, many CMARL algorithms can be applied. Specifically, we take QMIX~\cite{qmix} as the problem solver, 
where there exists a Q network $Q_i(\tau^i,a^i)$ for each agent $i$ and a mixing network that takes each Q value along with the global state as input and produces the value of $Q_{tot}$. Under the Dec-POMDP $\mathcal{\tilde M}$ proposed in Thm.~\ref{thm3}, we parameterize QMIX with $\theta$ and train it through minimizing:
\begin{equation}
    \begin{aligned}
        L_{ego}(\theta) = \mathbb{E}[(Q_{tot}(\boldsymbol{\tilde{\tau}, a}, \tilde{s};\theta)-y_{tot})^2],
    \end{aligned}
    \label{loss_qmix}
\end{equation}
where $y_{tot}=\tilde r+\gamma\max_{\boldsymbol{a'}} Q_{tot}(\boldsymbol{\tilde{\tau}', a'}, \tilde{s}';\theta^{-})$, and $\theta^{-}$ are parameters of a periodically updated target network.
 
In our ROMANCE framework,  we select a population of adversarial attackers from the archive, alternatively optimize the adversarial attackers or ego-system by fixing the other, and update the archive accordingly. The full algorithm of our ROMANCE can be seen in Algo.~2 in Appendix.

\section{Experiments}

In this section, we conduct extensive experiments to answer the following questions: 1) Can ROMANCE\footnote{Code is available at https://github.com/zzq-bot/ROMANCE} achieve high robustness compared to other baselines in different scenarios? 2) Can ROMANCE obtain a set of attackers with high attacking quality and diversity? 3) Can ROMANCE be integrated into multiple CMARL methods, and how does each hyperparameter influence the performance of ROMANCE?

We conduct experiments on SMAC~\cite{smac}, a widely used combat scenario of StarCraft II unit micromanagement tasks, where we train the ally units to beat enemy units controlled by the built-in AI with an unknown strategy. At each timestep, agents can move or attack any enemies and receive a global reward equal to the total damage done to enemy units. Here we consider multiple maps include maps 2s3z, 3m, 3s\_vs\_3z, 8m, MMM, and 1c3s5z. The detailed descriptions are presented in Appendix.

To ensure fair evaluation, we carry out all the experiments with five random seeds, and the results are presented with a 95\% confidence interval. Detailed network architecture, hyperparameter setting of ROMANCE are shown in Appendix.

\begin{figure}[ht!]
  \centering
  \includegraphics[scale=0.29]{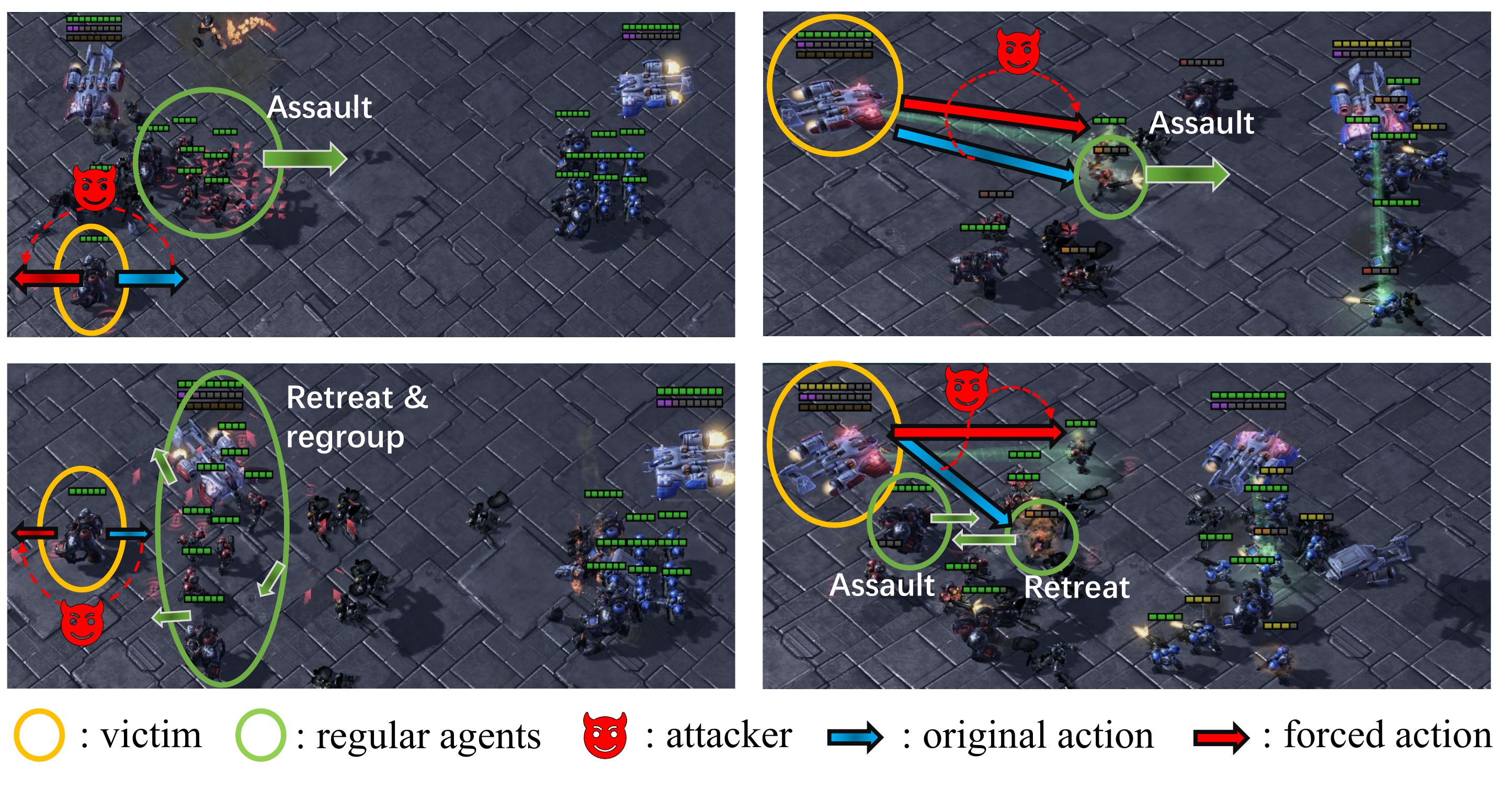}
   	\caption{Visualization of action selection under unseen action attack. The first and second rows show the policy learned without and with ROMANCE, respectively.}
	\label{visualization}
\end{figure}

\subsection{Competitive Results and Analysis} \label{robustness}
We implement ROMANCE based on QMIX~\cite{qmix} for its widely proven coordination ability. Then, ROMANCE is compared against four baselines: the vanilla QMIX, which is obtained to complete the task without any adversarial training, RANDOM, which adds random attack during training, and two strong baselines named RARL~\cite{pinto2017robust} and RAP~\cite{vinitsky2020robust}.

\textbf{RARL}~\cite{pinto2017robust} trains a robust agent in the presence of an adversarial attacker who applies disturbance to the system. Iteratively, the attacker learns to be an optimal destabilization policy, and then the agent learns to fulfill the original task while being robust to the adversarial attacker.

\textbf{RAP}~\cite{vinitsky2020robust} extends RARL by introducing population based training. At each rollout, it samples an attacker uniformly from the population and trains the agent and attackers iteratively as RARL does. The introduction of population improves the generalization by forcing the agent to be robust to a wide range of attackers, thus avoiding being exploited by some specific attackers.

\subsubsection{Robustness Visualization}

At first glance, we conduct experiments on map MMM to investigate how the training framework influences the coordination policy behavior under unpredictable attacks. As shown in Fig.~\ref{visualization}, when one coordinator from a well-trained coordination policy suffers from an action attack, the policy without adversarial training will still take the original coordination pattern but ignore the emergent situation, resulting in a sub-optimal policy. As seen at the beginning of the battle, the survivors learned by vanilla QMIX still try to assault but ignore the attacked Marauder when
it is drawn away by a malicious attack, causing severe damage to the team. Still, our ROMANCE can obtain a policy where the survivors retreat and wait for the Marauder to regroup for efficient coordination. At the middle stage of the battle, when some Marines are close to death but cannot get healed because the Medivac's action is being attacked, the survivors with full health learned by ROMANCE will charge forward to replace the dying Marine and cover him to let him get healed, while policy learned by vanilla QMIX still ramble in the map but ignore the teammates.
  
\begin{table*}[h!]
\centering
\resizebox{0.95\textwidth}{!}{
\begin{tabular}{|c|c|c|c|c|c|c|c|c|} 
\hline
\multicolumn{2}{|c|}{\diagbox{Method}{Map\_Name}} 
& \begin{tabular}[c]{@{}c@{}}2s3z \\$K=8$\end{tabular} & \begin{tabular}[c]{@{}c@{}}3m \\$K=4$\end{tabular} 
& \begin{tabular}[c]{@{}c@{}}3s\_vs\_3z\\$K=8$\end{tabular} 
& \begin{tabular}[c]{@{}c@{}}8m\\$K=5$\end{tabular}  
& \begin{tabular}[c]{@{}c@{}}MMM\\$K=8$\end{tabular} 
& \begin{tabular}[c]{@{}c@{}}1c3s5z\\$K=6$\end{tabular}
& \multicolumn{1}{c|}{$+/-/\approx$} 
\\ 
\hline
\multirow{6}{*}{Natural}       & vanilla QMIX                                              & $92.8\pm1.62$                                       & $\mathbf{97.9\pm1.02}$                                     & $98.3\pm0.78$                                            & $98.2\pm0.45$                                     & $95.8\pm1.59$                                      & $88.8\pm2.13$  & $1/1/4$                                                            \\
                               & RARL                                                      & $96.4\pm1.19$                                       & $86.0\pm5.38$                                     & $80.6\pm27.5$                                           & $95.3\pm3.31$                                     & $89.3\pm7.01$                                      & $76.9\pm9.85$       & $0/4/2$                                                       \\
                               & RAP                                                       & $\mathbf{98.1\pm0.76}$                                       & $91.3\pm4.93$                                     & $\mathbf{99.3\pm0.51}$                                            & $91.7\pm7.96$                                     & $95.3\pm4.98$                                      & $86.7\pm10.5$                                     & $0/1/5$                        \\
                               & RANDOM                                                    & $98.0\pm0.60$                                       & $95.3\pm2.07$                                     & $99.6\pm0.35$                                            & $\mathbf{98.6\pm0.90}$                                     & $93.8\pm7.56$                                      & $93.1\pm4.41$                                              & $1/0/5$                \\
                               &  {ROMANCE}               &  $97.9\pm1.34$   &  $96.0\pm1.83$ &  $97.8\pm1.78$        &  $94.3\pm3.94$ &  $\mathbf{97.1\pm1.49}$  &  $\mathbf{93.9\pm1.24}$   
                               & \\
\hline
\multirow{6}{*}{\begin{tabular}[c]{@{}c@{}}Random \\Attack\end{tabular}} & vanilla QMIX                                              & $78.8\pm1.28$                                       & $\mathbf{78.7\pm1.49}$                                     & $87.0\pm0.36$                                            & $66.2\pm2.08$                                     & $70.0\pm3.97$                                      & $66.6\pm2.03$                                      & $0/5/1$                        \\
                               & RARL                                                      & $84.3\pm2.40$                                       & $67.6\pm5.01$                                     & $70.1\pm29.1$                                           & $75.7\pm7.00$                                     & $62.2\pm10.2$                                     & $56.5\pm10.8$          &   $0/5/1$                                                \\
                               & RAP                                                       & $87.3\pm1.87$                                       & $73.5\pm3.49$                                     & $89.8\pm4.81$                                            & $\mathbf{78.4\pm8.22}$                                     & $84.2\pm9.05$                                      & $66.8\pm9.66$            &   $0/1/5$                                                  \\
                               & RANDOM                                                    & $83.9\pm6.38$                                       & $76.4\pm2.27$                                     & $91.9\pm1.32$                                            & $72.0\pm3.46$                                     & $72.9\pm7.09$                                      & $60.5\pm21.3$              &   $0/2/4$                                               \\
                               &  {ROMANCE}               &  $\mathbf{89.1\pm1.97}$   &  $78.1\pm5.13$ &  $\mathbf{93.0\pm1.82}$        &  $76.2\pm5.36$ &  $\mathbf{85.8\pm8.66}$  &  $\mathbf{77.9\pm1.96}$  &                         \\
\hline
\multirow{6}{*}{EGA}    & vanilla QMIX                                              & $26.7\pm4.28$                                       & $20.7\pm2.13$                                     & $30.9\pm1.52$                                            & $42.7\pm9.79$                                     & $37.9\pm3.13$                                      & $35.2\pm8.66$                                   &   $0/6/0$                           \\
                               & RARL                                                      & $56.1\pm11.8$                                      & $86.1\pm0.98$                                     & $60.9\pm14.2$                                           & $66.3\pm7.25$                                     & $41.5\pm11.6$                                     & $35.3\pm4.00$                                       &   $0/6/0$                       \\
                               & RAP                                                       & $64.1\pm11.9$                                      & $84.0\pm4.27$                                     & $65.1\pm4.41$                                            & $84.4\pm8.88$                                     & $74.9\pm15.5$                                     & $45.4\pm6.83$                     &   $0/4/2$                                         \\
                               & RANDOM                                                    & $48.3\pm17.3$                                      & $66.2\pm16.6$                                    & $54.4\pm7.83$                                            & $55.6\pm12.5$                                    & $53.1\pm6.09$                                      & $43.3\pm10.3$           
                               &   $0/6/0$                                         \\
                               &  {ROMANCE}               &  $\mathbf{81.6\pm0.84}$   &  $\mathbf{89.7\pm1.52}$ &  $\mathbf{90.5\pm1.97}$        &  $\mathbf{86.2\pm5.11}$ &  $\mathbf{84.0\pm11.5}$ &  $\mathbf{66.5\pm3.24}$   
                               &\\
\hline
\end{tabular}}
\caption{Average test win rates of different methods under various attack settings, where $K$ is the number of attacks during training, ``Natural" means no attack during testing, ``Random Attack" indicates every agent in the ego-system may be attacked randomly, and ``EGA" means our evolutionary generation based attackers. The best result of each 
column is highlighted in bold. The symbols `$+$, `$-$' and `$\approx$' indicate that the result is significantly superior to, inferior to, and almost equivalent to ROMANCE, respectively, according to the Wilcoxon rank-sum test~\cite{mann1947test} with confidence level $0.05$.}\label{mainrobustness}
\end{table*}


\begin{table*}
\centering
\resizebox{0.95\textwidth}{!}{
\begin{tabular}{c|c|c|c|c|c|c|c|c} 
\hline
Method      & $K=6$            & $K=7$            &  \cellcolor[rgb]{0.893,0.893,0.893} $K=8$            & $K=9$             & $K=10$            & $K=11$            & $K=12$           & $K=14$            \\ 
\hline
vanilla QMIX                              & $59.2\pm2.66$ & $42.1\pm0.81$ &   \cellcolor[rgb]{0.893,0.893,0.893}$26.7\pm4.28$ & $17.3\pm0.62$ & $12.2\pm0.33$ & $8.74\pm0.14$ & $6.42\pm0.84$ & $2.82\pm0.70$  \\ 
\hline
RARL                                      & $72.7\pm4.22$ & $65.2\pm9.11$ &   \cellcolor[rgb]{0.893,0.893,0.893}$56.1\pm11.8$ & $46.3\pm12.1$  & $38.0\pm13.5$  & $31.8\pm13.1$  & $25.9\pm12.3$ & $18.6\pm10.9$  \\ 
\hline
RAP                                       & $81.7\pm7.37$ & $73.6\pm7.46$ &   \cellcolor[rgb]{0.893,0.893,0.893}$64.1\pm11.9$ & $53.5\pm11.7$  & $42.5\pm11.6$  & $33.9\pm11.4$  & $25.8\pm10.7$ & $14.0\pm7.50$  \\ 
\hline
RANDOM                                    & $69.3\pm10.9$ & $56.8\pm12.8$ &   \cellcolor[rgb]{0.893,0.893,0.893}$48.3\pm17.3$ & $34.7\pm17.3$  & $25.5\pm15.8$  & $19.8\pm14.3$  & $14.9\pm12.6$ & $10.0\pm9.69$  \\ 
\hline
{ROMANCE} & $\mathbf{89.9\pm1.19}$ & $\mathbf{86.4\pm1.87}$ &   \cellcolor[rgb]{0.893,0.893,0.893}$\mathbf{81.6\pm0.84}$ & $\mathbf{75.1\pm0.58}$ & $\mathbf{66.7\pm1.56}$  & $\mathbf{57.4\pm1.61}$  & $\mathbf{48.6\pm2.60}$ & $\mathbf{41.5\pm2.17}$  \\
\hline
\end{tabular}}
\caption{Average test win rates of each method when the test number of attacks $K$ changes on map 2s3z. The best result of each column is highlighted in bold, and the column for the training number (i.e., $K=8$) is highlighted as gray.} \label{generalizationtest}
\end{table*}

\subsubsection{Training phase evaluation}
\begin{figure}[htbp]
\centering
    \begin{minipage}{0.47\linewidth}
	\centering
	\includegraphics[width=1\linewidth]{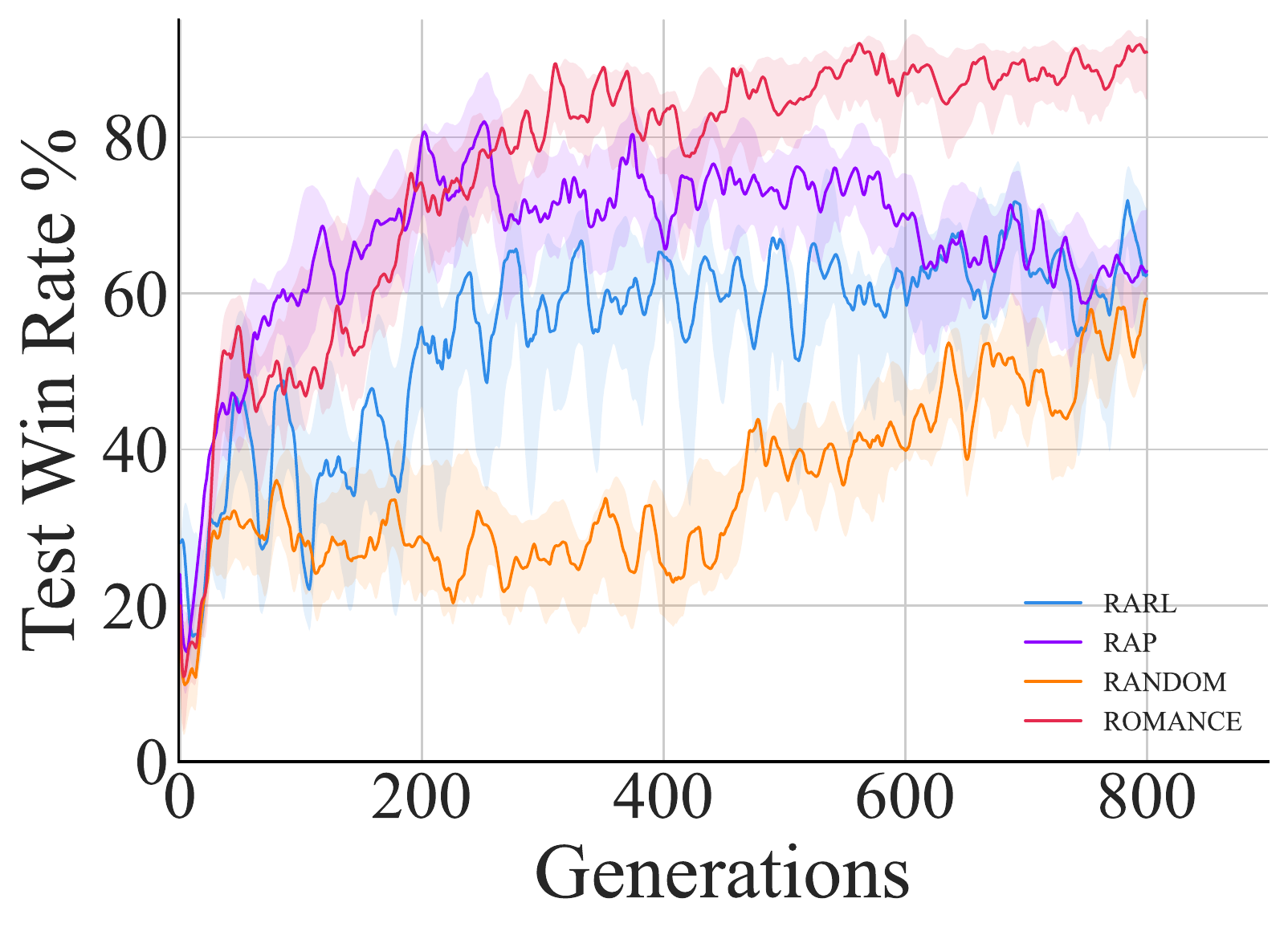}
    \centerline{(a) 3s\_vs\_3z}
    \end{minipage}
	\begin{minipage}{0.47\linewidth}
	\centering
	\includegraphics[width=1\linewidth]{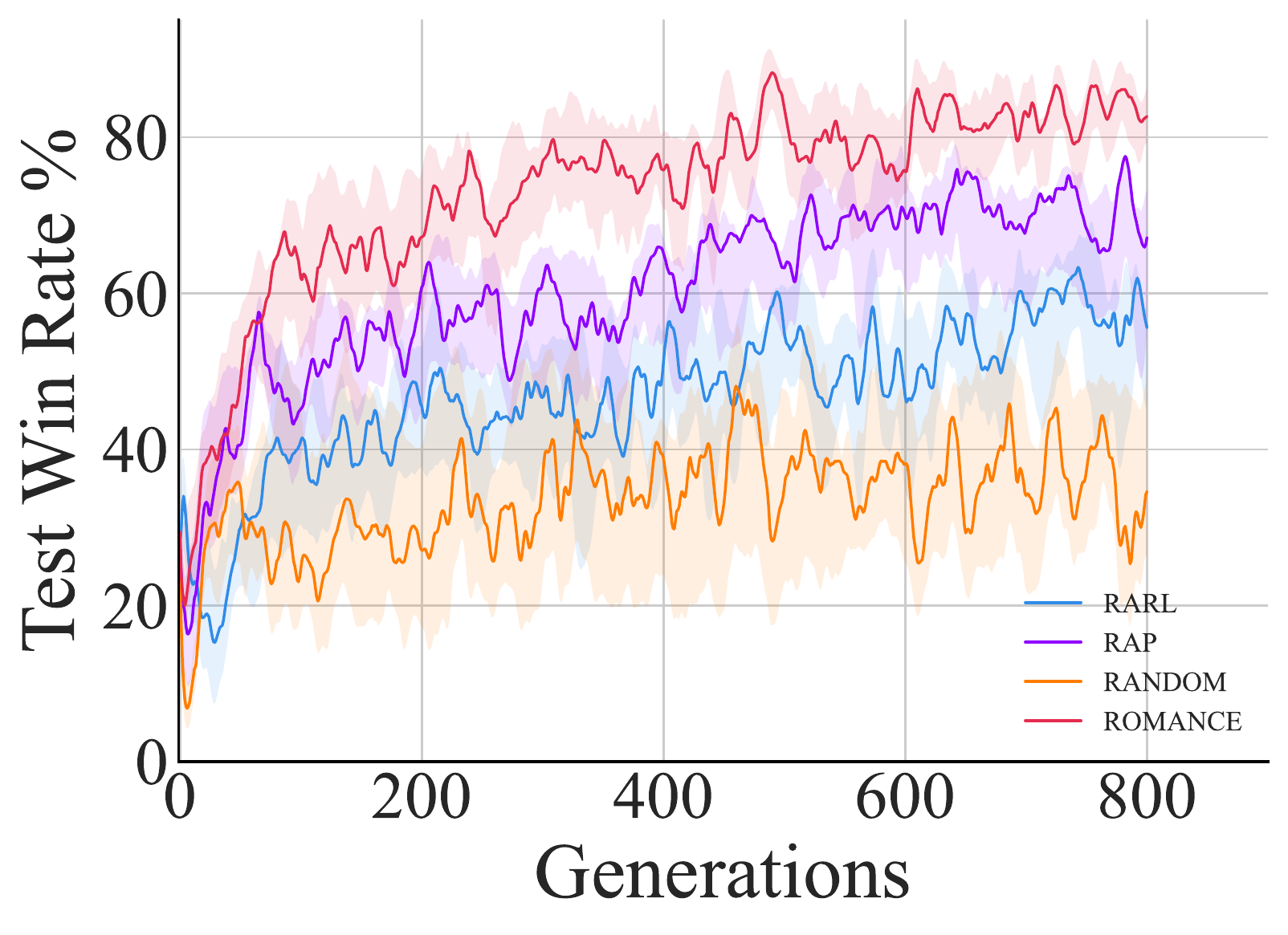}
	\centerline{(b) 2s3z}
    \end{minipage}
\caption{Averaged test win rates on two maps. }
\label{trainingrecord}
\end{figure}
Fig.~\ref{trainingrecord} shows the learning curves of different methods of maps 3s\_vs\_3z and 2s3z when facing fixed unknown attackers. ROMANCE outperforms all baselines in both maps at each generation either in terms of convergence speed or asymptotic performance. 
RANDOM achieves the worst performance in almost every generation, indicating that adding random policy perturbation can somehow improve the exploration ability under an environment without attack but has no effect on tasks where attackers exist. The superiority of ROMANCE over RARL and RAP demonstrates the necessity of adversarial population and diverse population training, respectively.
In Fig.~\ref{vdn_qplex_curve}, we present the learning curves of different methods implemented on QPLEX and VDN on map 2s3z. The curves show that ROMANCE can
significantly enhance the robustness of value-based MARL algorithms when they are integrated.

\begin{figure}[h!]
    \centering
	    \begin{minipage}{0.492\linewidth}
		\centering
		\includegraphics[width=1\linewidth]{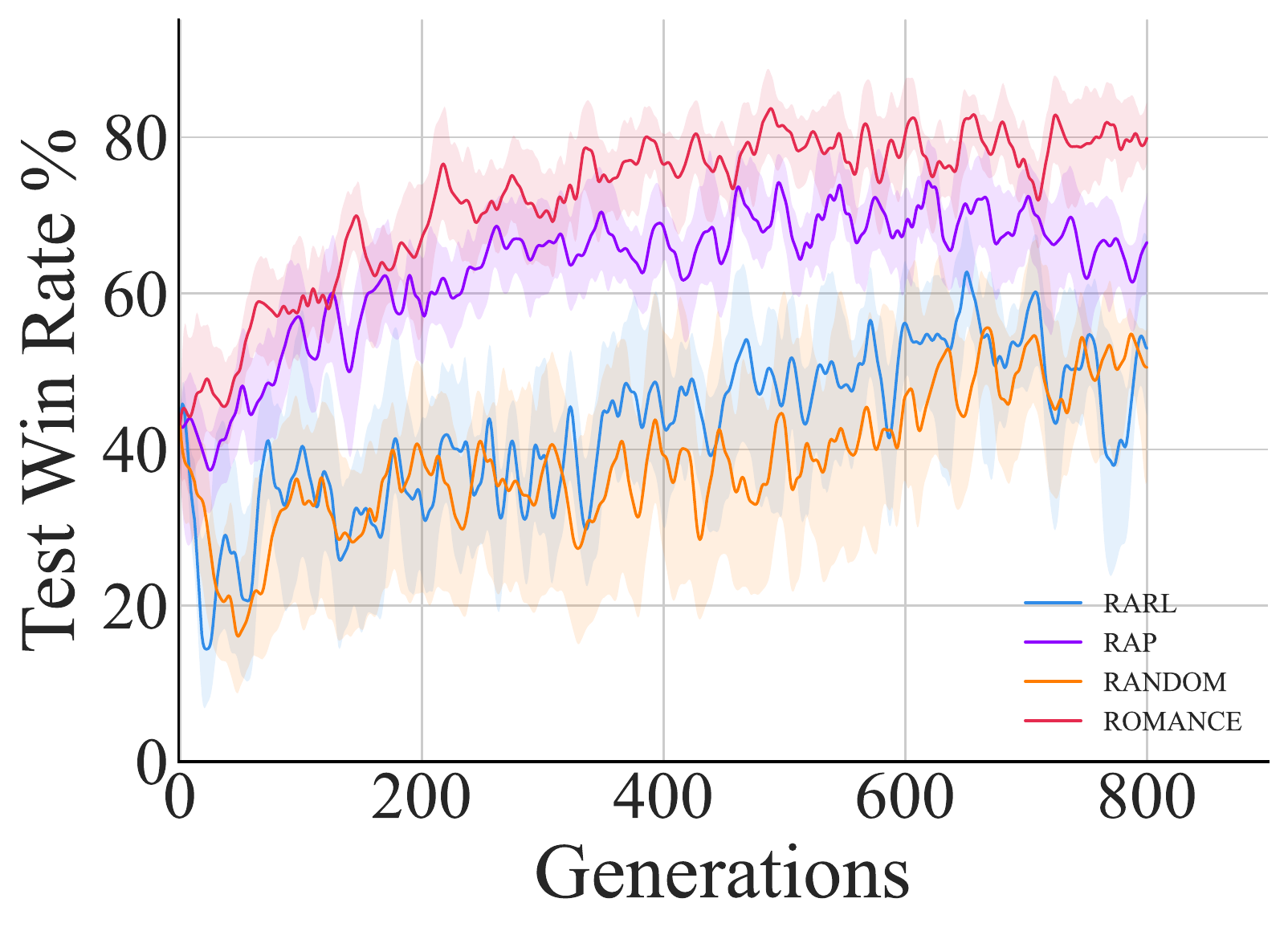}
		\centerline{(a) VDN}
	    \end{minipage}
	    \begin{minipage}{0.492\linewidth}
		\centering
		\includegraphics[width=1\linewidth]{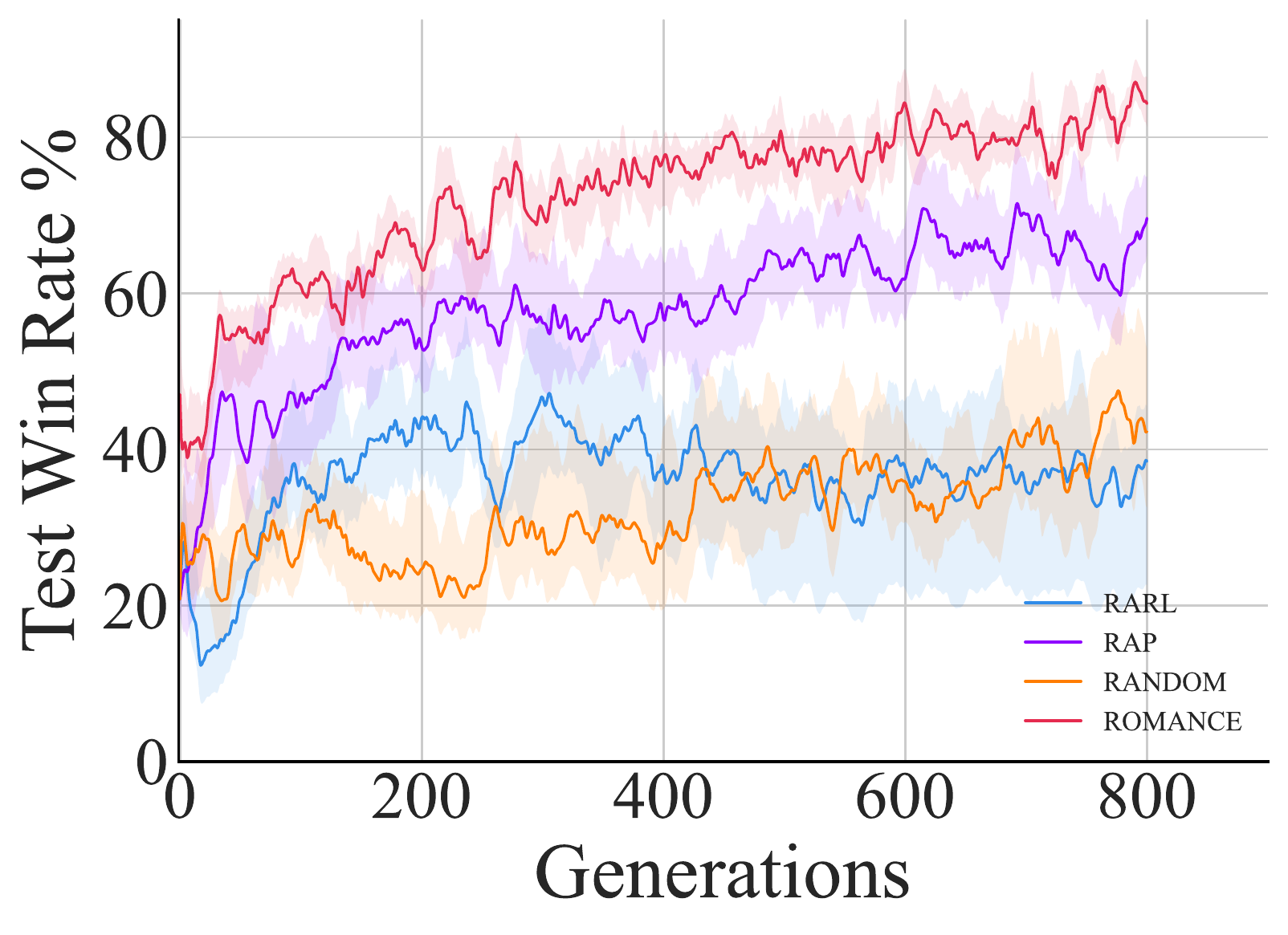}
		\centerline{(b) QPLEX}
	    \end{minipage}
	\caption{Average test win rates of VDN and QPLEX on map 2s3z during the training phase.}
    \label{vdn_qplex_curve}
\end{figure}

\subsubsection{Robustness Comparison}
We here come to show whether ROMANCE can improve the coordination ability under different unpredictable attacks compared with multiple baselines. As shown in Tab.~\ref{mainrobustness},  we present three settings, where ``Natural" means no attackers or the attack number $K=0$ during testing, ``Random Attack" indicates that every agent in the ego-system might be attacked randomly at each step, 
``EGA (Evolutionary Generation based Attackers)" are unseen attackers with high performance and diversity generated by our method in multiple extra runs, which could be viewed as out-of-distribution attackers for different methods. In the ``Natural'' setting, ROMANCE achieves comparable or better performance compared to other baselines. RARL achieves inferiority over other baselines because it aims to learn a worst-case performance, leading to a pessimistic result for the coordination ability. RAP and RANDOM show superiority over the vanilla QMIX in some maps, such as 2s3z. We believe this is because random attacks or a weak adversarial population during training can promote exploration for MARL under natural circumstances.
Furthermore, when suffering from a random attack during testing (i.e., the ``Random Attack'' setting), vanilla QMIX has the most remarkable performance decrease in most maps, demonstrating the multi-agent coordination policy's vulnerability without any adversarial training. Methods based on adversarial training such as RANDOM, RARL, and RAP show superiority over vanilla QMIX, indicating that adversarial training can improve the robustness of MARL policy. 
We further find that when encountering strong attackers (i.e., the ``EGA'' setting), all baselines sustain a severe performance decrease. The proposed ROMANCE achieves a high superiority over other baselines on most maps under different attack modes, indicating it can indeed learn a robust coordination policy under different policy perturbation conditions.

\subsubsection{Beyond-limited-budget evaluation}
As this study considers a setting where the number of attacks is fixed during the training phase, we evaluate the generalization ability when altering the attack budget during testing. We conduct experiments on map 2s3z with the number of attacks $K=8$ during training. As shown in Tab.~\ref{generalizationtest}, when the budget is different from the training phase, policy learned with vanilla QMIX and RANDOM sustain a severe performance decrease even when the budget slightly changes, indicating that these two methods may overfit to the training situation and lack of generalization. RARL and RAP show superiority over vanilla QMIX and RANDOM, demonstrating that adversarial training can relieve the overfitting problem but is still inferior to ROMANCE along with the budget increasing, manifesting the high generalization ability gained by the training paradigm of ROMANCE.

\subsection{Attacker Population Validation}
\label{attackerpopulation}

\begin{figure*}[htbp]
  \centering
  \includegraphics[scale=0.41]
  {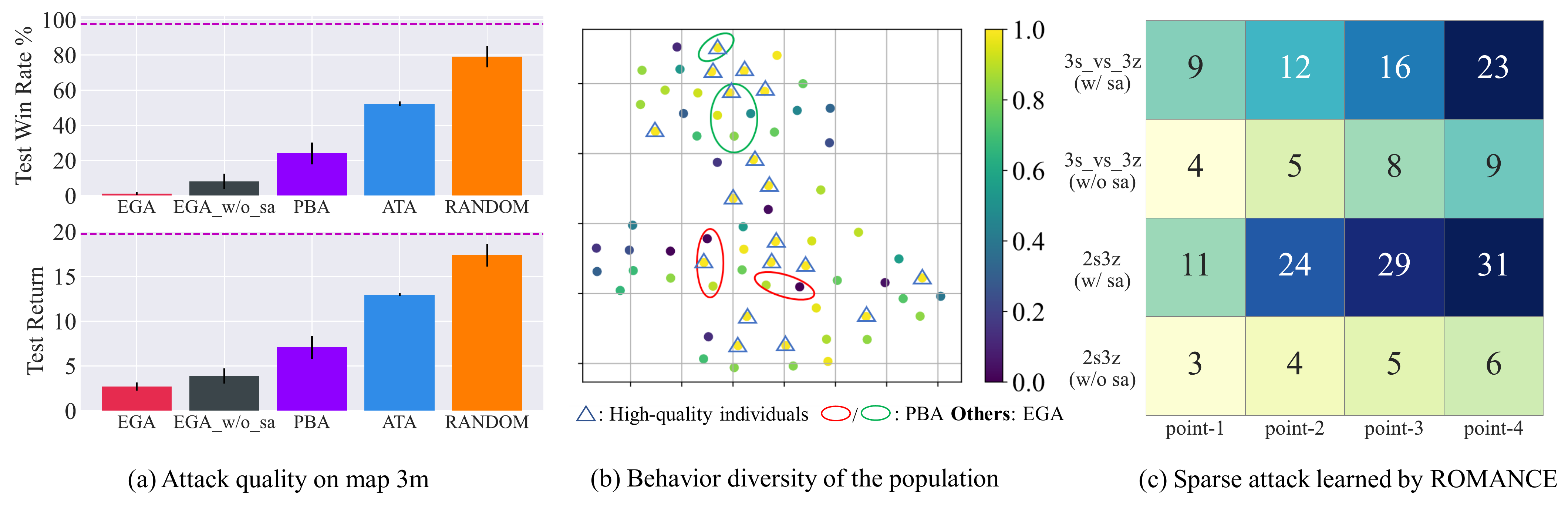}
  \caption{Attacker quality validation, where EGA and EGA\_w/o\_sa are our Evolutionary Generation of Attackers with and without sparse action regularizer, respectively; PBA and ATA refer to Population-Based Attackers and Alternate Training Attackers, respectively; RANDOM means that we select a coordinator to attack randomly; (a) The attacking quality. (b) The t-SNE projection of attackers on map 3m. (c) Attack points produced on two maps with and without sparse action regularizer.} 
  \label{attackerquality}
\end{figure*} 

As our method needs to maintain a set of adversarial attackers, we design experiments to investigate the attackers generated by our  method. As shown in Fig.~\ref{attackerquality}(a), the multiple attack methods can correspondingly decrease the ego-system's performance, and our EGA (Evolutionary Generation based Attackers) show high superiority over others both in return and test win rate, demonstrating the effectiveness of the proposed training paradigm. EGA, EGA\_w/o\_sa, and PBA (Population-Based Attackers) outperform ATA (Alternating Training Attackers) and RANDOM, indicating that population training can indeed improve the attacking ability. Nevertheless, PBA works inefficiently, showing that only randomly initialized individuals in the population are insufficient for efficient population training. The superiority of EGA over its ablation EGA\_w/o\_sa demonstrates the effectiveness of sparse action regularizer.

Furthermore, we analyze the behavior representations learned by each attacking method in a two-dimensional
plane using the t-SNE method~\cite{van2008visualizing}. As shown in Fig.~\ref{attackerquality}(b), we can discover that the traditional population-based training paradigm can only generate attackers with very limited diversity and quality, with most points gathering around in a few specific areas, while our EGA can find widespread points with both high quality and diversity. Fig.~\ref{attackerquality}(c) shows that our sparse action regularizer can efficiently promote the sparse attacking points to disperse as much as possible within one episode, which can also implicitly promote the diversity of the attackers in a population by preventing attackers from exhausting the attack opportunities at the very beginning.

\subsection{Integrative and Parameter Sensitive Studies} \label{moreresult}
\begin{figure}
\centering
\subfigure[Integrative Abilities]{
\label{Integrative}
\includegraphics[width=0.22\textwidth]{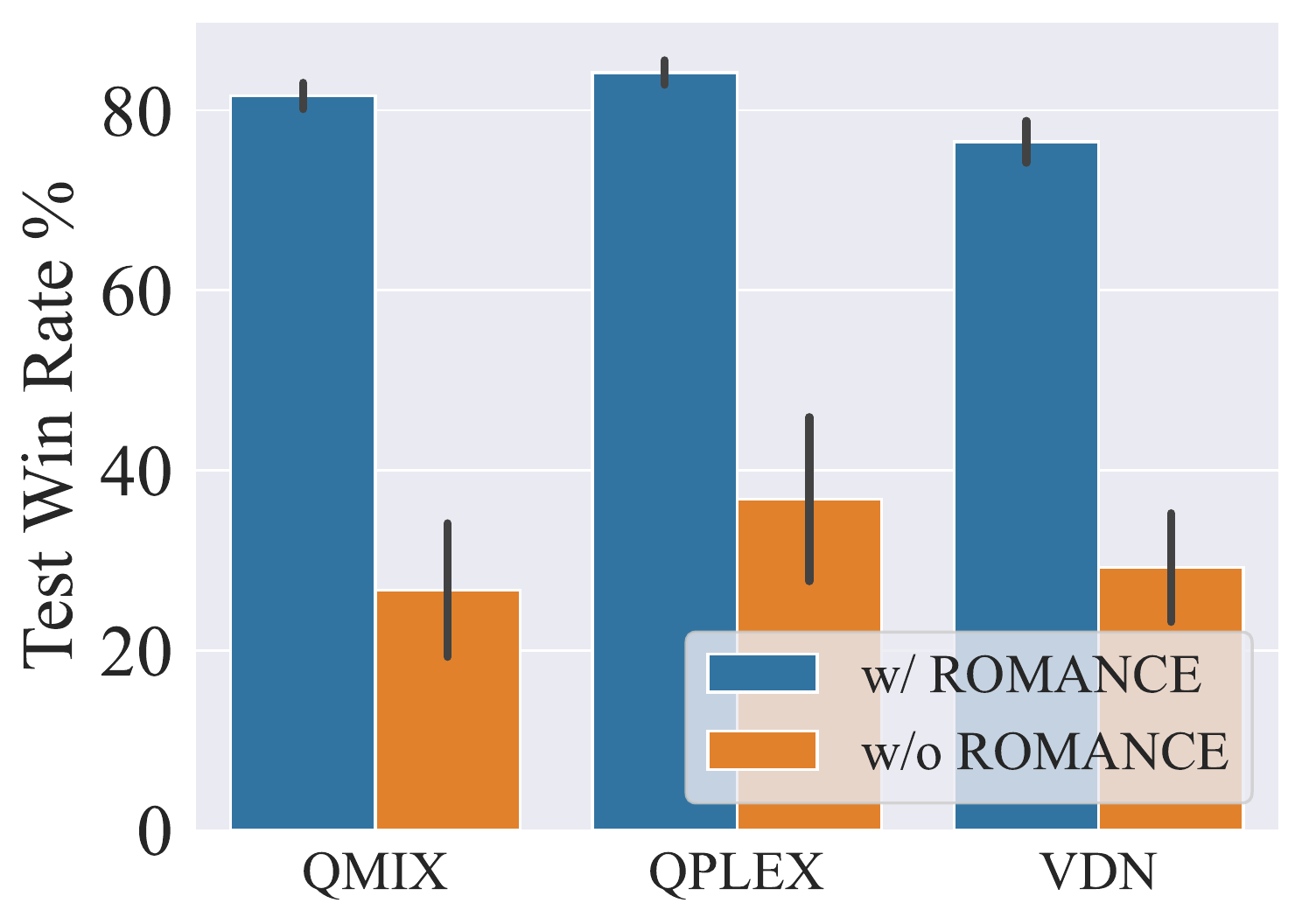}
}
\subfigure[Sensibility of $\alpha$]{
\label{Sensitive}
\includegraphics[width=0.22\textwidth]{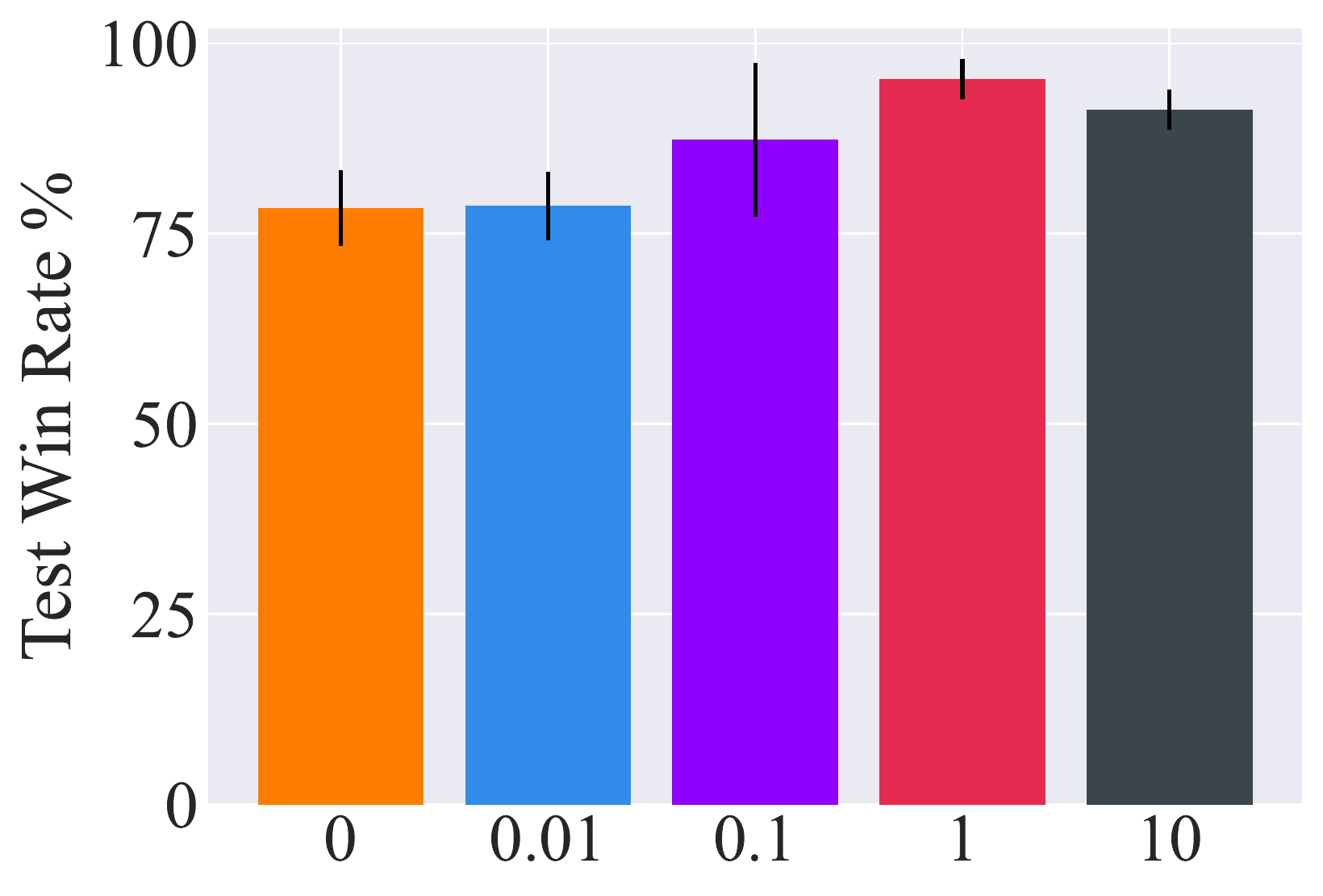}
}

\caption{More experiments about ROMANCE. }
\label{ablation}
\end{figure}

ROMANCE is agnostic to specific value decomposition MARL methods. We can regard it as a plug-in model and integrate it with existing MARL value decomposition methods like QPLEX~\cite{qplex},  QMIX~\cite{qmix}, and VDN~\cite{vdn}. As shown in Fig.~\ref{Integrative}, when integrating with ROMANCE, the performance of the baselines vastly improves on map 2s3z, indicating that the proposed training paradigm can significantly enhance robustness for these value-based MARL methods.
 
One of the crucial elements of our framework is the hyper-parameter $\alpha$ which controls the attack quality and diversity. We here conduct experiments to study the sensibility of $\alpha$ in Eq.~(\ref{loss_tot}) for the whole framework. As shown in Fig.~\ref{Sensitive}, on map 3s\_vs\_3z, the performance is more influenced when we set $\alpha = 0$ or $0.01$, which refers to optimizing the attacking quality only but almost ignoring the diversity goal, indicating the importance of behavior diversity for the population training. Nevertheless, other choices have a negligible impact on the performance. We believe this is because the customized selection and update operators of the evolution mechanism can help balance the two goals for a slightly larger $\alpha$. More experimental results, such as how each parameter influences ROMANCE, are shown in Appendix.

\section{Conclusion}
This paper considers the robust cooperative MARL problem, where some coordinators suffer from unpredictable policy perturbation. We first formalize this problem as an LPA-Dec-POMDP, where some coordinators from a team may sustain action perturbation accidentally and unpredictably. We then propose ROMANCE, an efficient approach to learn robust multi-agent coordination via evolutionary generation of auxiliary adversarial attackers. Experimental results on robustness and generalization testing verify the effectiveness of ROMANCE, and more analysis results also confirm it from 
multiple aspects. As our method aims to learn a disentangled adversarial attacker policy, which demands a heuristic-based policy perturbation function, future work on more reasonable and efficient ways such as observation perturbation and automatic search for the best budget for different tasks would be of great value. Furthermore, how to design efficient and effective robust multi-agent reinforcement learning algorithms for the open-environment setting~\cite{zhou2022open} is also valuable for the MARL community.   

\section*{Acknowledgments}
 This work is supported by National Key Research and Development Program of China (2020AAA0107200), the National Science Foundation of China (61921006, 62022039), and the program B for Outstanding Ph.D. candidate of Nanjing University. We would like to thank Jingcheng Pang, Chengxing Jia, and the anonymous reviewers for their helpful discussions and support.

\section{Proofs for LPA-Dec-POMDP}
\begin{myLemma}[Bellman equation for fixed $\boldsymbol{\pi}$ and $\pi_{adv}$] Given $\hat {\mathcal{M}} = \langle\mathcal{N,S,A}, P, K, \Omega,$ $O,R,\gamma \rangle$, $\boldsymbol{\pi}$ and $\pi_{adv}$, we have 
\begin{equation}
    \begin{aligned}
        \hat V_{\boldsymbol{\pi}\circ \pi_{adv}}(s,k) 
        =& \mathbb{E}_{\boldsymbol{a,\hat{a}}\sim\boldsymbol{\pi}\circ \pi_{adv}(s,k)}[\mathbb{E}_{s'\sim{P(\cdot|s',\boldsymbol{a})}}[R(s, \boldsymbol{\hat a}, s')\\
        +&\gamma  \hat V_{\boldsymbol{\pi}\circ \pi_{adv}}(s',k-\mathbb{I}(\boldsymbol{\hat a}\neq \boldsymbol{a}))]],
    \end{aligned}
\end{equation}
where $\boldsymbol{\pi}(\boldsymbol{a}|s) = \prod_{i\in\mathcal{N}} \pi^i(a^i|O(s, i))$,  $\mathbb{E}_{\boldsymbol{a,\hat{a}}\sim\boldsymbol{\pi}\circ \pi_{adv}(s,k)}$ $[f]=\sum_{\boldsymbol{a}\in\mathcal{A}} \boldsymbol{\pi}(\boldsymbol{a}|s)\sum_{\boldsymbol{\hat a}\in\mathcal{A}} \pi_{adv}(\boldsymbol{\hat{a}}|s,\boldsymbol{a},k) f$. 
\begin{proof}\renewcommand{\qedsymbol}{}
    Based on the definition of $\hat V_{\boldsymbol{\pi}\circ \pi_{adv}}(s,k)$:
    \begin{equation}
    \begin{aligned}
        &\hat V_{\boldsymbol{\pi}\circ \pi_{adv}}(s,k)\\ &=\mathbb{E}_{\boldsymbol{\pi}\circ \pi_{adv}}[\sum_{n=0}^{\infty} \gamma^{n} r_{t+n+1}|s_t=s,k_t=k]\\
        &=\mathbb{E}_{\boldsymbol{\pi}\circ \pi_{adv}}[r_{t+1}+\sum_{n=0}^{\infty} \gamma^{n} r_{t+n+2}|s_t=s,k_t=k]\\
        &=\sum_{\boldsymbol{a}\in\mathcal{A}} \boldsymbol{\pi}(\boldsymbol{a}|s)\sum_{\boldsymbol{\hat a}\in\mathcal{A}} \pi_{adv}(\boldsymbol{\hat{a}}|s,\boldsymbol{a},k)\sum_{s'\in\mathcal{S}} P(s'|s,\boldsymbol{\hat a})[r_{t+1}+\\
        &\gamma \mathbb{E}_{\boldsymbol{\pi}\circ \pi_{adv}}[\sum_{n=0}^{\infty} \gamma^{n} r_{t+n+2}|s_{t+1}=s',k_{t+1}=k-\mathbb{I}(\boldsymbol{\hat a}
        \neq \boldsymbol{a})]]\\
        &= \mathbb{E}_{\boldsymbol{a,\hat{a}}\sim\boldsymbol{\pi}\circ \pi_{adv}(s,k)}[\mathbb{E}_{s'\sim{P(\cdot|s',\boldsymbol{\hat a})}}[R(s, \boldsymbol{\hat a}, s')\\
        &+\gamma  \hat V_{\boldsymbol{\pi}\circ \pi_{adv}}(s',k-\mathbb{I}(\boldsymbol{\hat a}\neq \boldsymbol{a}))]].
    \end{aligned}
    \label{eq2}
    \end{equation}
    The state-value function $\hat Q_{\boldsymbol{\pi\circ \pi_{adv}}}(s,\boldsymbol{a},k)$ can be derived in a similar way.
\end{proof}

\end{myLemma}
\begin{myThm} 
    Given an LPA-Dec-POMDP $\hat {\mathcal{M}} = \langle\mathcal{N,S,A},$ $ P, K, \Omega,O, R,\gamma \rangle$, a fixed joint policy $\boldsymbol{\pi}$ of the ego-system and a heuristic-based policy perturbation function $g$, there exists an MDP $\bar{\mathcal{M}}=(\mathcal{\bar{S},\bar{A}},\bar{P}, \bar R, \gamma)$ such that the optimal adversarial attacker $\pi_{adv}^*$ for LPA-Dec-POMDP is disentangled by an optimal policy $v^*$ of $\bar{\mathcal{M}}$ and $g$, where $\mathcal{\bar S}=\mathcal{S}\times \mathbb{N}$, $\bar s=(s, k), \bar s'=(s', k'), k, k' \le K$ indicates the remaining attack budget, $\mathcal{\bar A}=\mathcal N\cup \{null\}$, $\bar R(\bar s,\bar a,\bar s')=-R(s, \boldsymbol{\hat{a}},s')$,
    \begin{equation*}
        \begin{aligned}
            \bar P(\bar s'|\bar s, \bar a)=&\begin{cases}
         0\qquad\qquad\qquad\qquad\,\, k-k'\notin \{0,1\}\\
        P(s'|s,\boldsymbol{\hat a})\mathbb{I}(\boldsymbol{\hat a}=\boldsymbol{a})\quad k-k'=0\\
        P(s'|s,\boldsymbol{\hat a})\mathbb{I}(\boldsymbol{\hat a}\neq \boldsymbol{a})\quad k-k'=1
            \end{cases},
        \end{aligned}
    \end{equation*}
    \label{thm1}
   where $\bar d(\bar s_0)=d(s_0)\mathbb{I}(k_0=K)$,  $d$ and $\bar d$ are distributions over initial state in $\hat{\mathcal{M}}$ and $\bar{\mathcal{M}}$, respectively, $\boldsymbol{a} = \boldsymbol{\pi}(s)$, $\boldsymbol{\hat a} = g(\bar a,\boldsymbol{\pi}(s),k)$ are original and forced action of ego-system, respectively.
\end{myThm}
\begin{proof}\renewcommand{\qedsymbol}{}
    To prove this theorem, we first prove that, given a fixed joint ego-agent policy $\boldsymbol{\pi}$ and a heuristic-based policy perturbation function $g$, $\forall v$, we have $\hat V_{\boldsymbol{\pi}\circ \pi_{adv}}(s,k) = -\bar V_{v}(\bar s)$, where $\pi_{adv}=v\circ g$, $\bar s = (s,k)$.
    
    Based on the Bellman equation for $\bar V$, we could derive:
    \begin{equation}
        \begin{aligned}
            &\bar V(\bar s) \\
            &= \sum_{\bar a\sim \mathcal{\bar A}} v(\bar a|\bar s)\sum_{\bar{s}'\sim\mathcal{\bar S}} P(\bar s'|\bar s, \bar a)[\bar R(\bar s,\bar a,\bar s')+\gamma \bar V(\bar s')]\\
            &=\sum_{i\in\mathcal{\hat N}}v(i|s,k)[\sum_{s'\sim \mathcal S} P(s'|s,\boldsymbol{\hat a})\mathbb{I}(\boldsymbol{\hat a}= \boldsymbol{v}(s))[-R(s, \boldsymbol{\hat{a}},s')+\\
            &\quad\gamma \bar V_{v}((s',k))]+\sum_{s'\sim \mathcal S} P(s'|s,\boldsymbol{\hat a})\mathbb{I}(\boldsymbol{\hat a}\neq \boldsymbol{v}(s))[-R(s, \boldsymbol{\hat{a}},s')\\
            &\quad-\gamma \bar V_{v}((s',k-1))]]\\
            &= \sum_{i\in\mathcal{\hat N}}v(i|s,k)[\sum_{s'\sim \mathcal S} P(s'|s,\boldsymbol{\hat a})[-R(s, \boldsymbol{\hat{a}},s')+\\
            &\quad\gamma \bar V_{v}((s',k-\mathbb{I}(\boldsymbol{\hat a}\neq \boldsymbol{a})))]\\
            &=\sum_{\boldsymbol{a}\in\mathcal{A}} \boldsymbol{\pi}(\boldsymbol{a}|s)\sum_{\boldsymbol{\hat a}\in\mathcal{A}} \pi_{adv}(\boldsymbol{\hat{a}}|s,\boldsymbol{a},k)\sum_{s'\in\mathcal{S}} P(s'|s,\boldsymbol{\hat a})[-R(s,\\&\quad \boldsymbol{\hat{a}},s')+
        \gamma \bar{V}((s',k-\mathbb{I}(\boldsymbol{\hat a}\neq \boldsymbol{a}))).
            \end{aligned}    
            \label{eq3}
    \end{equation}
    Then, we discuss the case that $\forall s\in \mathcal{S},k=0$, where $\pi_{adv}(\boldsymbol{\hat a}|s,\boldsymbol{a},0)=\mathbb{I}(\boldsymbol{\hat a}=\boldsymbol{a})$:
    \begin{equation}
        \begin{aligned}
            &\hat V_{\boldsymbol{\pi}\circ \pi_{adv}}(s,0)
        \\=& \sum_{\boldsymbol{a}\in\mathcal{A}} \boldsymbol{\pi}(\boldsymbol{a}|s)\sum_{s'\in\mathcal{S}} P(s'|s,\boldsymbol{a})[R(s,\boldsymbol{\hat a},s')+
        \gamma \hat{V}_{\boldsymbol{\pi}\circ\pi_{adv}}(s',0)]\\
        =& V_{\boldsymbol{\pi}}(s)\\
        &\bar V_v((s,0))
        \\=& \sum_{\bar a\sim\mathcal{\bar A}}v(\bar a|\bar s)g(\bar a,\boldsymbol{\pi}(s),0)\sum_{s'\sim \mathcal S} P(s'|s,\boldsymbol{\hat a})[-R(s, \boldsymbol{\hat{a}},s')
        +\\&\quad\gamma \bar V_{v}((s,0))]\\
        =&\sum_{\boldsymbol{a}\in\mathcal{A}} \boldsymbol{\pi}(\boldsymbol{a}|s)\sum_{s'\in\mathcal{S}} P(s'|s,\boldsymbol{a})[-R(s,\boldsymbol{\hat a},s')+\gamma \bar V_{v}((s,0))]\\
        =&-V_{\boldsymbol{\pi}}(s),
        \end{aligned}
    \end{equation}
    where $V_{\boldsymbol{\pi}}(s)$ is the value function under the original Dec-POMDP.
    
    Accordingly, we have that $k=0$, $\hat V_{\boldsymbol{\pi}\circ \pi_{adv}}(s,0) =-\bar V_v((s,0)), \forall s\in \mathcal{S}$. Then we could derive that $\hat V_{\boldsymbol{\pi}\circ \pi_{adv}}(s,k) =-\bar V_v((s,k)), , \forall s\in \mathcal{S}, \forall k\in \{0,1,...,K\}$ based on Eq.(\ref{eq2}) and Eq.(\ref{eq3}).
    
    Besides, for $\bar M$, we have $\forall(\bar s,\bar a,\bar s')$,
    \begin{equation*}
        -\max_{s,\boldsymbol{\hat {a}},s'} R(s,\boldsymbol{\hat a},s')\leq \bar R(\bar s,\bar a,\bar s')\leq -\min_{s,\boldsymbol{\hat {a}},s'} R(s,\boldsymbol{\hat a},s').
    \end{equation*}
\end{proof}
Based on the basic property of MDP, there exist an optimal policy $v^*$ for $\bar M$, such that $\bar V_{v^*}(\bar s)\geq \bar V_v(\bar s)$,$\forall \bar s\in \mathcal{\bar S}, \forall v$. 

Now we have that the optimal adversarial attacker $\pi_{adv}^*$ for LPA-Dec-POMDP is disentangled by an optimal policy $v^*$ of $\bar M$ and $g$, that is $\pi^*_{adv}=v^*\circ g$. By optimizing $v$ in $\mathcal{\bar M}$, we are able to optimize the adversarial attacker in $\mathcal{\hat M}$.

\begin{myThm}
     Given $\hat {\mathcal{M}} = \langle\mathcal{N,S,A}, P, K, \Omega,O, R,\gamma \rangle$, a fixed deterministic adversarial attacker policy $\pi_{adv}$, there exists a Dec-POMDP $\mathcal{\tilde{M}} = \langle \mathcal{N,\tilde{S},A},\tilde{P},\Omega,\tilde{O},\tilde{R},\gamma\rangle$ such that the optimal policy of $\mathcal{\tilde{M}}$ is the optimal policy for LPA-Dec-POMDP given $\pi_{adv}$, where $\mathcal{\tilde{S} }=\mathcal{S}\times\mathbb{N}$, $\tilde d(\tilde s_0)=d(s_0)\mathbb{I}(k_0=K)$,  $\tilde{O}(\tilde{s},i)=O(s,i)$, $\tilde R(\tilde s,\boldsymbol{a},\tilde s')=R(s,\boldsymbol{\hat{a}},s')$,
     \begin{equation*}
        \begin{aligned}
            \tilde P(\tilde {s}'| \tilde s,\boldsymbol{a})=
            \begin{cases}
            0 \qquad\qquad\quad\,\,\,\, k-k'\notin \{0,1\}\\
            P(s'|s,\boldsymbol{\hat a})\qquad otherwise
            \end{cases},
        \end{aligned}
    \end{equation*}
    where $\tilde d$ and $d$ are distributions over initial state in $\mathcal{\tilde M}$ and $\mathcal{\hat M}$, respectively, and $\tilde{s} = (s,k)$,  $\tilde{s}' = (s',k')$, $\boldsymbol{\hat a}=\pi_{adv}(s,\boldsymbol{a},k)$.
    \label{thm2}
\end{myThm}
\begin{proof}\renewcommand{\qedsymbol}{}
    Following the same idea in the proof of Thm.\ref{thm1}, we aim to prove that, given a deterministic $\pi_{adv}$, $\forall \boldsymbol{\pi}$,we have $\hat V_{\boldsymbol{\pi}\circ \pi_{adv}}(s,k) = \tilde V_{\boldsymbol{\pi}}(\tilde s)$, $\forall s\in\mathcal{S}, k\in\{0,1,...,K\}$, where $\tilde s = (s,k)$.
    
    For $\mathcal{\tilde M}$, we have Bellman Equation:
\begin{equation}
    \begin{aligned}
        &\tilde V_{\boldsymbol{\pi}}(\tilde s)\\
        &=\sum_{\boldsymbol{a}\sim \mathcal{A}} \boldsymbol{\pi}(\boldsymbol{a}|\tilde s)\sum_{\tilde s'\in\mathcal{\tilde S}}\tilde P(\tilde{s}'|\tilde s,\boldsymbol{a})[\tilde R(\tilde s,\boldsymbol{a},\tilde s')+\gamma \tilde V_{\boldsymbol{\pi}}(\tilde s')]\\
        &= \sum_{\boldsymbol a\sim \mathcal{A}} \boldsymbol{\pi}(\boldsymbol{a}|\tilde s)\sum_{s'\in\mathcal S} P(s'|s,\boldsymbol{\hat a})[R(s,\boldsymbol{\hat a},s')+\gamma\tilde V_{\boldsymbol{\pi}}((s', k\\&-\mathbb{I}(\boldsymbol{\hat a}\neq \boldsymbol{ a})))],
    \end{aligned}
    \label{eq5}
\end{equation}
where $\boldsymbol{\hat a}=\pi_{adv}(s,\boldsymbol{a},k)$.

By substituting $\pi_{adv}(\boldsymbol{\hat a}|s,\boldsymbol{a},k)$ with a deterministic policy in  Eq.(\ref{eq2}), we have:

\begin{equation}
    \begin{aligned}
        &\hat V_{\boldsymbol{\pi}\circ \pi_{adv}}(s,k)\\
        &=\sum_{\boldsymbol a\sim \mathcal{A}} \boldsymbol{\pi}(\boldsymbol{a}|s)\sum_{s'\in\mathcal S} P(s'|s,\boldsymbol{\hat a})[R(s,\boldsymbol{\hat a})+\gamma\hat V_{\boldsymbol{\pi}\circ \pi_{adv}}((s', k\\&-\mathbb{I}(\boldsymbol{\hat a}\neq \boldsymbol{ a})))].
    \end{aligned}
    \label{eq6}
\end{equation}

We do not distinguish between $\boldsymbol{\pi}(\boldsymbol{a}|\tilde s)$ and $\boldsymbol{\pi}(\boldsymbol{a}|s)$ since $\boldsymbol{\pi}(\boldsymbol{a}|\tilde s)=\prod_{i\in\mathcal N}\pi^i(a^i|\tilde O(\tilde s,i))=\prod_{i\in\mathcal N}\pi^i(a^i|O(s,i))=\boldsymbol{\pi}(\boldsymbol{a}|s)$.

Similar to the proof in Thm.(\ref{thm1}), we could derive that $\forall \boldsymbol{\pi}$, $k=0$, we have $\hat V_{\boldsymbol{\pi}\circ \pi_{adv}}(s,0) = \tilde V_{\boldsymbol{\pi}}(\tilde s)$, $\forall s\in\mathcal{S}$, where $\tilde s = (s,0)$. Combined with Eq.(\ref{eq5}) and Eq.(\ref{eq6}),for $\forall \boldsymbol{\pi}$, we have $\hat V_{\boldsymbol{\pi}\circ \pi_{adv}}(s,k) = \tilde V_{\boldsymbol{\pi}}(\tilde s)$, $\forall s\in\mathcal{S}, k\in\{0,1,...,K\}$, where $\tilde s = (s,k)$. The optimal policy $\boldsymbol{\pi}$ in $\tilde M$ is also the optimal policy in original LPA-Dec-POMDP $\mathcal{\hat M}$.
\end{proof}

\begin{myThm}
     Given $\hat {\mathcal{M}} = \langle\mathcal{N,S,A}, P, K, \Omega,O, R,\gamma \rangle$, a stochastic adversarial attacker policy $\pi_{adv}$, there exists an Dec-POMDP $\mathcal{\tilde{M}} = \langle \mathcal{N,\tilde{S},A},\tilde{P},\Omega,\tilde{O},\tilde{R},\gamma\rangle$ such that $\forall \boldsymbol{\pi}$, we have $\tilde{V}_{\boldsymbol{\pi}}(\tilde s)\leq \hat V_{\boldsymbol{\pi}\circ\pi_{adv}}(s,k)$, where $\tilde s=(s,k)$, $\hat V_{\boldsymbol{\pi}\circ\pi_{adv}}(s,k)$ denotes the state value function in the original LPA-Dec-POMDP, $\forall s\in \mathcal S, \forall k \in \{0,1,...,K\}$.
\end{myThm}
\begin{proof}\renewcommand{\qedsymbol}{}
    Similar to Thm.(\ref{thm2}), we define the reward and transition functions as follows:
    \begin{equation*}
        \begin{aligned}
            &\tilde R(\tilde{s},\boldsymbol{a},\tilde{s}') = \begin{cases}
                R(s,\boldsymbol{a},s')\qquad\qquad\qquad\quad k-k'=0\\
                \hat R(s,\boldsymbol{a},s',k)\qquad\qquad\qquad otherwise
            \end{cases}\\
            &\tilde P(\tilde {s}'| \tilde s,\boldsymbol{a})=
            \begin{cases}
            0 \qquad\qquad\qquad\qquad\quad\quad\,\,\,\, k-k'\notin \{0,1\}\\
            P(s'|s,\boldsymbol{a})\pi_{adv}(\boldsymbol{a}| s,\boldsymbol{a},k)\quad k-k'=0\\
           \hat P(s'|s,\boldsymbol{a},k)\qquad\qquad\qquad\, k-k'=1
           \end{cases},
        \end{aligned}
    \end{equation*}
    where $\hat P(s'| s,\boldsymbol{a}, k)=\sum_{\boldsymbol{\hat a}\sim\mathcal{A},\boldsymbol{\hat a}\neq \boldsymbol{a}}P(s'| s,\boldsymbol{\hat a})$, $\hat R(s,\boldsymbol{a},s')=$ $\sum_{\boldsymbol{\hat a}\sim\mathcal{A},\boldsymbol{\hat a}\neq \boldsymbol{a}} R(s,\boldsymbol{\hat a}, s')$.
    
    By substituting $\pi_{adv}$ with a stochastic version in Eq.(\ref{eq5}), we have 
    \begin{equation}
        \begin{aligned}
        &\tilde{V}_{\boldsymbol{\pi}}(\tilde s)
        \\&= \sum_{\boldsymbol a\sim \mathcal{A}} \boldsymbol{\pi}(\boldsymbol{a}|\tilde s)\pi_{adv}(\boldsymbol{a}|s,\boldsymbol{a},k)\sum_{s'\in\mathcal S} P(s'|s,\boldsymbol{ a})[R(s,\boldsymbol{ a},s')\\
        &+\gamma\tilde V_{\boldsymbol{\pi}}((s', k))]+\sum_{\boldsymbol a\sim \mathcal{A}} \boldsymbol{\pi}(\boldsymbol{a}|\tilde s)\sum_{s'\in\mathcal S}\sum_{\boldsymbol{\hat a}\neq \boldsymbol{a}}\pi_{adv}(\boldsymbol{\hat a}|s,\boldsymbol{a},k)\\
        &\, P(s'|s,\boldsymbol{\hat a})[\sum_{\boldsymbol{\hat a}'\neq \boldsymbol{a}}\pi_{adv}(\boldsymbol{\hat a}'|s,\boldsymbol{a},k)R(s,\boldsymbol{\hat a}',s')
        +\\&\quad\gamma\tilde V_{\boldsymbol{\pi}}((s', k-1))].
        \end{aligned}
    \end{equation}
    By splitting Eq.(\ref{eq2}) based on $\mathbb{I}(\boldsymbol{\hat a}\neq \boldsymbol{a})$, we would derive that:
    \begin{equation}
        \begin{aligned}
        &\hat{V}_{\boldsymbol{\pi}\circ\pi_{adv}}(s,k)
        \\&= \sum_{\boldsymbol a\sim \mathcal{A}} \boldsymbol{\pi}(\boldsymbol{a}|\tilde s)\pi_{adv}(\boldsymbol{a}|s,\boldsymbol{a},k)\sum_{s'\in\mathcal S} P(s'|s,\boldsymbol{ a})[R(s,\boldsymbol{ a},s')\\
        &+\gamma\hat V_{\boldsymbol{\pi}\circ\pi_{adv}}(s', k)]+\sum_{\boldsymbol a\sim \mathcal{A}} \boldsymbol{\pi}(\boldsymbol{a}|\tilde s)\sum_{s'\in\mathcal S}\sum_{\boldsymbol{\hat a}\neq \boldsymbol{a}}\pi_{adv}(\boldsymbol{\hat a}|s,\boldsymbol{a},k)\\
        &\, P(s'|s,\boldsymbol{\hat a})[R(s,\boldsymbol{\hat a},s')
        +\gamma\hat V_{\boldsymbol{\pi}\circ \pi_{adv}}(s', k-1)].
        \end{aligned}
    \end{equation}
    
    Similar to the proof in Thm.(\ref{thm1}), we could derive that $\forall \boldsymbol{\pi}$, $k=0$, and we have $\hat V_{\boldsymbol{\pi}\circ \pi_{adv}}(s,0) = \tilde V_{\boldsymbol{\pi}}(\tilde s)$, $\forall s\in\mathcal{S}$, where $\tilde s = (s,0)$.
    
    According to the basic property of expectation that $\mathbb{E}_{X}[f(X)g(X)]\geq \mathbb{E}_{X}[f(X)]\mathbb{E}_X[g(X)]$, we have  $\forall \boldsymbol{\pi}$, and $\hat V_{\boldsymbol{\pi}\circ \pi_{adv}}(s,k) \geq \tilde V_{\boldsymbol{\pi}}(\tilde s)$, $\forall s\in\mathcal{S}, k\in\{0,1,...,K\}$, where $\tilde s = (s,k)$. By optimizing $\boldsymbol{\pi}$ in Dec-POMDP $\mathcal{\tilde M}$, we are optimizing the lower bound of the state value function of $\boldsymbol{\pi}$ in the original LPA-Dec-POMDP.
\end{proof}

\section{Additional Description of Algorithms}
\subsection {Description of related algorithms}
\label{algorithm}
As described in the main manuscript, our work mainly includes two parts: Attacker Population Generation and the whole process of ROMANCE, the detailed pseudocodes are shown in Algo.~\ref{algo1} and Algo.~\ref{algo2}, respectively.

We describe the procedure of evolutionary generation of the attacker population in Algorithm~\ref{algo1}. In each iteration, we first select $n_p$ adversarial attackers from the archive as the current population based on their quality scores. Ego-system will interact with the attacker in the population alternately as is described in the function CollectTraj. The trajectories collected are used to optimize the population and thus obtaining $n_p$ new adversarial attackers. To keep the attackers' quality and diversity in the archive, the specialized updating mechanism in function Update is adopted.
 \begin{algorithm}[t!]
 \caption{Evolutionary Generation of Attackers}\label{algo1}
 \SetKwProg{Fn}{Function}{:}{\KwRet $Arc_{adv}$}
 \SetKwProg{aFn}{Function}{:}{\KwRet $\mathcal{D_{\boldsymbol{\pi}}}$,$\mathcal{D}_{adv}^j$}
 \SetKwFunction{FMain}{Update}
 \SetKwFunction{FMainx}{CollectTraj}
 \KwIn{A joint ego-agent policy $\boldsymbol{\pi}$, archive $Arc_{adv}$, population size $n_p$, archive capacity $n_a$.}
 $P_{adv}\leftarrow $select$(Arc_{adv},n_p)$\;
 \For{$\pi_{adv}^{\phi_j} \in P_{adv}$}{
     $\mathcal{D_{\boldsymbol{\pi}}}, \mathcal{D}_{adv}^j\leftarrow$ CollectTraj($\boldsymbol{\pi},\pi_{adv}^{\phi_j}$)\;
 }
 Optimize $\{\phi_j\}_{j=1}^{n_p}$ based on Eq.(6)\;
 $Arc_{adv}\leftarrow$ Update($Arc_{adv}, P_{adv}$)\;
 \Fn{\FMain(\text{Arc}$_{adv}$, \text{P}$_{adv}$)}{
 \For{$\pi_{adv}^{\phi_{j}} \in P_{adv}$}{
     \For{$\pi_{adv}^{\phi_{i}} \in Arc_{adv}$}{calculate $\text{Dist}(i,j)$ based on Eq.(8)\;}
    
     \eIf{$\min_{i}\text{Dist}(i,j)\geq threshold$}
     {add($\pi_{adv}^{\phi_{j}}, Arc_{adv}$)\;}
     {reserve($\pi_{adv}^{\arg\min_{i}\text{Dist}(i,j)}, \pi_{adv}^j, Arc_{adv}$)\;}
 }}
 \aFn{\FMainx($\boldsymbol{\pi}$,$\pi_{adv}^{\phi_j}$) }{
 $k\leftarrow K$\;
 $\mathcal{D_{\boldsymbol{\pi}}}, \mathcal{D}_{adv}^j\leftarrow\{\},\{\}$\;
\For{$t=0$ \KwTo $T$}{
    $\{o^i_t\}_{i=1}^N= \{O(s_t,i)\}_{i=1}^N$\;
    $\boldsymbol{a}_t= \boldsymbol{\pi}(\boldsymbol{\tau}_t)$\;
    $\hat i\sim v^j(s_t,k)$\;
    \eIf{$\hat i\neq null$ and $k>0$}{
    $\hat{a}^{\hat i}_t=\arg\min_{a^{\hat i}} Q^i(\tau_t^{\hat i}, a^{\hat i})$\;
    $\boldsymbol{\hat a}_t = (\hat{a}^{\hat i}_t, \boldsymbol{a}_t^{-\hat i})$\;
    $k\leftarrow k-1$\;
    }{
    $\boldsymbol{\hat a}_t=\boldsymbol{a}_t$
    }
    $s_{t+1},r_{t},done\leftarrow env.step(\boldsymbol{\hat a}_t)$\;
    $\mathcal {D}_{\pi}\leftarrow\mathcal {D}_{\pi}\cup\{s_t, \boldsymbol{o}_t, \boldsymbol{a}_t, r_t,done\}$\;
     $\mathcal {D}_{adv}^j\leftarrow\mathcal {D}_{adv}^j\cup\{s_t, \hat{i}, -r_t,done\}$\;
     \If{done is True}{
        break;
     }
}
 }
\end{algorithm}

In Algorithm~\ref{algo2}, we present the alternating training paradigm ROMANCE, where the best response to a population of adversarial attackers are optimized based on the evolutionary generation of attackers in Algorithm~\ref{algo1}.

As for the two subject baselines, RARL and RAP, we present the detailed pseudocodes in Algo.~\ref{algo3} and Algo.\ref{algo4}, respectively.

\begin{algorithm}[t!]
\caption{ROMANCE}\label{algo2}
\KwIn{ Environment $\mathcal{E}$, population size $n_p$, archive capacity $n_a$, num of iterations $N_{gen}$.}
Initialize the archive $Arc_{adv}$ with capacity $n_a$\;
\For{$gen=1$ \KwTo $N_{gen}$}{$P_{adv}\leftarrow$ select$(Arc_{adv},n_p)$\;
     \For{$n=1$ \KwTo $N_{adv}$}{
        \For{$\pi_{adv}^{\phi_j} \in P_{adv}$}{$\mathcal{D_{\boldsymbol{\pi}}}, \mathcal{D}_{adv}^j\leftarrow$ CollectTraj($\boldsymbol{\pi}^{\theta},\pi_{adv}^{\phi_j}$)\;}
        Optimize $\boldsymbol{\phi}$ with $\mathcal{D}_{adv}$ based on Eq.~(6)\;
     }
 \For{$n=1$ \KwTo $N_{ego}$}{
 \For{$\pi_{adv}^{\phi_j} \in P_{adv}$}{$\mathcal{D_{\boldsymbol{\pi}}}, \mathcal{D}_{adv}^j\leftarrow$ CollectTraj($\boldsymbol{\pi}^{\theta},\pi_{adv}^{\phi_j}$)\;}
 Optimize $\theta$ with $\mathcal D_{\boldsymbol{\pi}}$ based on Eq.~(9)\;}
 
$Arc_{adv}\leftarrow$ Update($Arc_{adv}, P_{adv}$)\; 
 }
\end{algorithm}

\begin{algorithm}[t!]
\caption{RARL}\label{algo3}
\KwIn{ Environment $\mathcal{E}$, num of iterations $N_{gen}$.}
Initialize an attacker policy $\pi_{adv}^{\phi}$ \;
\For{$gen=1$ \KwTo $N_{gen}$}{
     \For{$n=1$ \KwTo $N_{adv}$}{
         $\mathcal D_{\boldsymbol{\pi}}, \mathcal{D}_{adv}^j\leftarrow$ CollectTraj($\boldsymbol{\pi}^{\theta},\pi_{adv}^{\phi_j}$)\;
         Optimize $\boldsymbol{\phi}$ with $\mathcal{D}_{adv}$ based on Eq.~(3)\;
         }
     
 \For{$n=1$ \KwTo $N_{ego}$}{
 $\mathcal{D_{\boldsymbol{\pi}}}, \mathcal{D}_{adv}^j\leftarrow$ CollectTraj($\boldsymbol{\pi}^{\theta},\pi_{adv}^{\phi_j}$)\;
 Optimize $\theta$ with $\mathcal D_{\boldsymbol{\pi}}$ based on Eq.~(9)\;}
 }
\end{algorithm}

\begin{algorithm}[h!]
\caption{RAP}\label{algo4}
\KwIn{ Environment $\mathcal{E}$, population size $n_p$, num of iterations $N_{gen}$.}
Initialize the population $P_{adv}$ with capacity $n_p$\;
\For{$gen=1$ \KwTo $N_{gen}$}{
     \For{$n=1$ \KwTo $N_{adv}$}{
        \For{$\pi_{adv}^{\phi_j} \in P_{adv}$}{$\mathcal{D_{\boldsymbol{\pi}}}, \mathcal{D}_{adv}^j\leftarrow$ CollectTraj($\boldsymbol{\pi}^{\theta},\pi_{adv}^{\phi_j}$)\;}
        Optimize $\boldsymbol{\phi}$ with $\mathcal{D}_{adv}$ based on Eq.~(3)\;
     }
 \For{$n=1$ \KwTo $N_{ego}$}{
 \For{$\pi_{adv}^{\phi_j} \in P_{adv}$}{$\mathcal{D_{\boldsymbol{\pi}}}, \mathcal{D}_{adv}^j\leftarrow$ CollectTraj($\boldsymbol{\pi}^{\theta},\pi_{adv}^{\phi_j}$)\;}
 Optimize $\theta$ with $\mathcal D_{\boldsymbol{\pi}}$ based on Eq.~(9)\;}
 }
\end{algorithm}

\section {Detailed description of SMAC}
\label{smac}
\begin{figure*}[htbp!]
    \centering
    \includegraphics[scale=0.5]{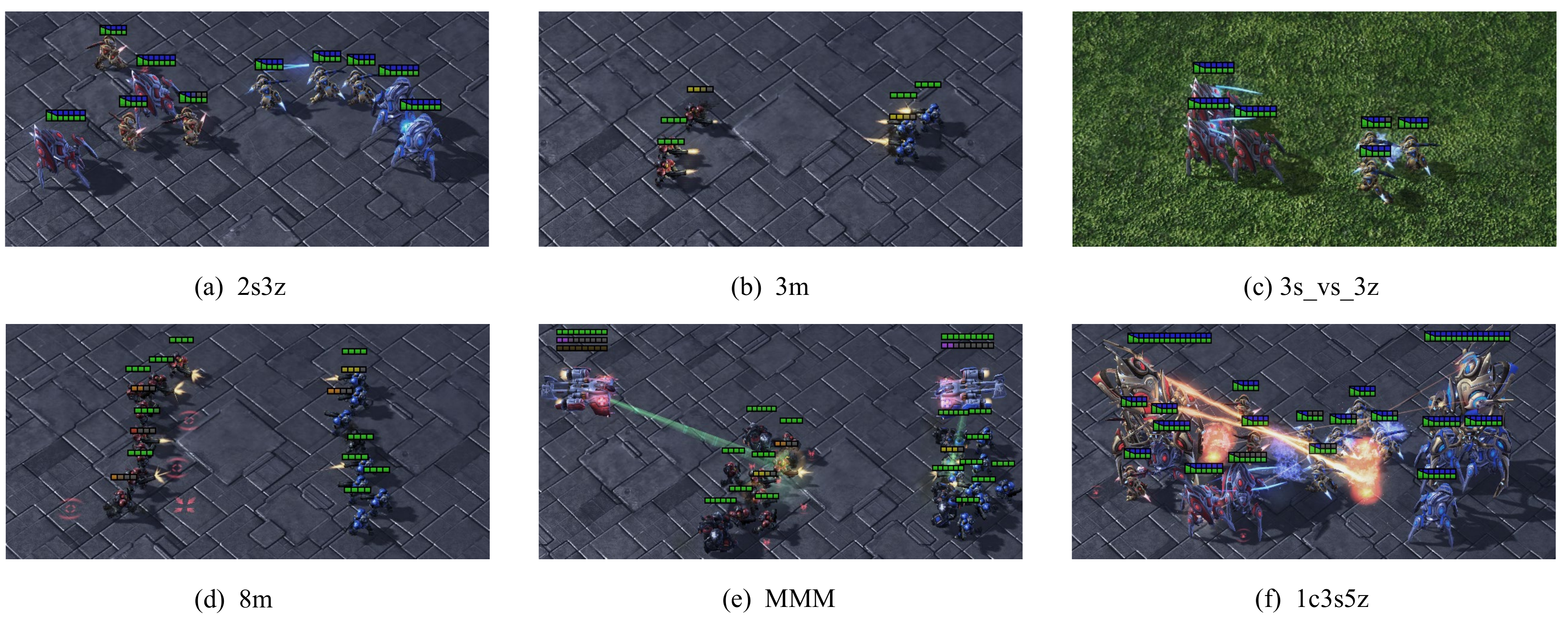}
    \caption{Snapshots of our selected StarCraft II scenarios.}
    \label{map_snap_shot}
\end{figure*}

SMAC~\cite{smac} is a combat scenario of StarCraft II unit micromanagement tasks. 
We consider a partial observation setting, where an agent can only see a circular area around it with a radius equal to the sight range, which is set to $9$. We train the ally units with reinforcement learning algorithms to beat enemy units controlled by the built-in AI. At the beginning of each episode, allies and enemies are generated at specific regions on the map. Every agent takes action from the discrete action space at each timestep, including the following actions: no-op, move [direction], attack [enemy id], and stop. Under the control of these actions, agents can move and attack in continuous maps. MARL agents will get a global reward equal to the total damage done to enemy units at each timestep. Killing each enemy unit and winning the combat (killing all the enemies) will bring additional bonuses of $10$ and $200$, respectively. We briefly introduce the SMAC maps used in our paper in Tab.~\ref{SMACmaps}, and the snapshots of each map are shown in Figure~\ref{map_snap_shot}.

\section{The Architecture, Infrastructure, and Hyperparameters Choices of ROMANCE}

Since ROMANCE is built on top of QMIX in the main experiments (VDN and QPLEX in the Integrative Abilities part in the main manuscript), we here present specific settings, including network architectures and hyperparameters choices. The local agent network shares the same architecture with QMIX, having a GRU cell with a dimension of 64 to encode historical information and two fully connected layers to compute local Q values. Mixing networks are applied according to existing MARL methods.
The adversarial attacker utilizes a multi-layer perceptron (MLP) with a hidden layer of $64$ units as a victim selection function to choose the victim and force it to execute the local worst action according to the heuristic-based policy perturbation function. We adopt RMSProp as the optimizer with $\alpha=0.99$, $\epsilon=1\times 10^{-5}$ for both ego-system and attackers. Specifically, the learning rate of the ego-system is set to be $4\times 10^{-4}$ for 2s3z, 3m, and 3s\_vs\_3z and $2\times 10^{-4}$ for others. $\delta=5\times10^{-2}$ and $\lambda=4\times 10^{-2}$ are the parameters of reference distribution and regularized factor for SPRQ, respectively. Besides, we set a smoothing constant $b=2\times 10^{-2}$ over the action distribution in case of KL-divergence approaching infinity. The whole framework is trained end-to-end with collected episodic data on NVIDIA GeForce RTX 2080 Ti GPUs with 800 iterations (generations). 
\begin{table}[]
\centering
\resizebox{0.474\textwidth}{!}{
\begin{tabular}{|c|c|c|c|} 
\hline
Map   & Ally Units                                                                                                                             & Enemy Units                                                                                                                      & Type                                                                                                                                   \\ 
\hline
2s3z       & \begin{tabular}[c]{@{}c@{}}\textcolor[rgb]{0.141,0.161,0.184}{2 Stalkers,}\\\textcolor[rgb]{0.141,0.161,0.184}{3 Zealots}\end{tabular} & \begin{tabular}[c]{@{}c@{}}\textcolor[rgb]{0.141,0.161,0.184}{2 Stalkers,}\\\textcolor[rgb]{0.141,0.161,0.184}{3 Zealots}\end{tabular} & \begin{tabular}[c]{@{}c@{}}Symmetric,\\~Heterogeneous\end{tabular}                                                                     \\ 
\hline
3m         & \textcolor[rgb]{0.141,0.161,0.184}{3 Marines}                                                                                          & \textcolor[rgb]{0.141,0.161,0.184}{3 Marines}                                                                                    & \begin{tabular}[c]{@{}c@{}}Symmetric,\\Homogeneous\end{tabular}                                                                        \\ 
\hline
3s\_vs\_3z & \textcolor[rgb]{0.141,0.161,0.184}{3 Stalkers}                                                                                         & \textcolor[rgb]{0.141,0.161,0.184}{3 Zealots}                                                                                    & \begin{tabular}[c]{@{}c@{}}\textcolor[rgb]{0.141,0.161,0.184}{micro-trick,}\\\textcolor[rgb]{0.141,0.161,0.184}{~kiting}\end{tabular}  \\ 
\hline
8m         & 8 Marines                                                                                                                              & 8 Marines                                                                                                                        & \begin{tabular}[c]{@{}c@{}}Symmetric,\\Homogeneous\end{tabular}                                                                        \\ 
\hline
MMM        & \begin{tabular}[c]{@{}c@{}}1 Medivac, \\ 2 Marauders, \\ 7 Marines\end{tabular}                                                        & \begin{tabular}[c]{@{}c@{}}1 Medivac, \\ 2 Marauders, \\ 7 Marines\end{tabular}                                                  & \begin{tabular}[c]{@{}c@{}}Symmetric,\\Heterogeneous\end{tabular}                                                                      \\ 
\hline
1c3s5z     & \begin{tabular}[c]{@{}c@{}}1 Colossi, \\ 3 Stalkers, \\ 5 Zealots\end{tabular}                                                         & \begin{tabular}[c]{@{}c@{}}1 Colossi, \\ 3 Stalkers, \\ 5 Zealots\end{tabular}                                                   & \begin{tabular}[c]{@{}c@{}}Symmetric,\\Heterogeneous\end{tabular}                                                                      \\
\hline
\end{tabular}}
\caption{Properties of 6 conducted SMAC scenarios.}
\label{SMACmaps}
\end{table}

\section{Additional Experimental Results} 

\begin{figure}[h!]
    \centering
    	\begin{minipage}{0.492\linewidth}
		\centering
		\includegraphics[width=1\linewidth]{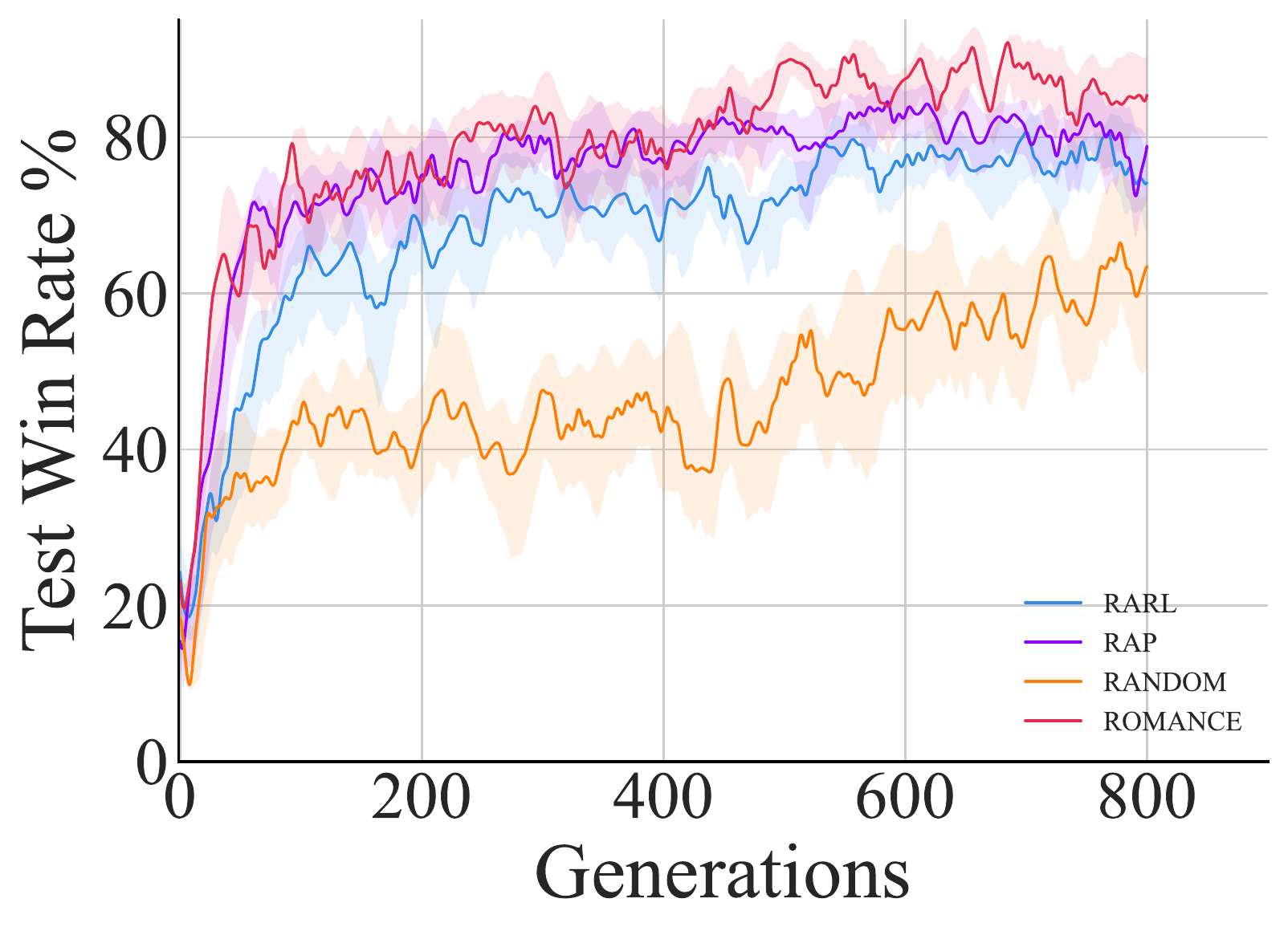}
		\centerline{(a) 3m}
	    \end{minipage}
	    \begin{minipage}{0.492\linewidth}
		\centering
		\includegraphics[width=1\linewidth]{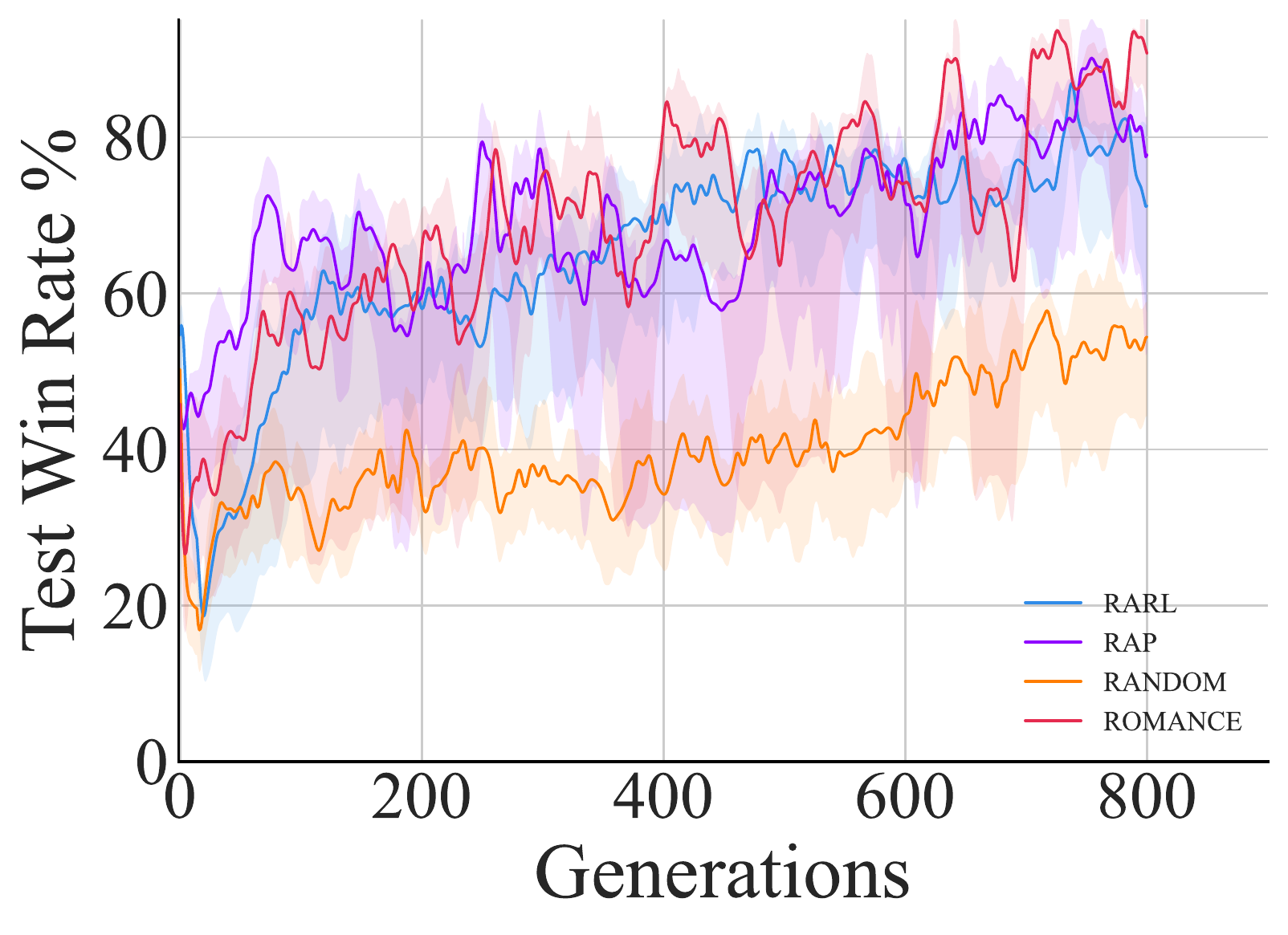}
		\centerline{(b) 8m}
	    \end{minipage}
	    
	    \begin{minipage}{0.492\linewidth}
		\centering
		\includegraphics[width=1\linewidth]{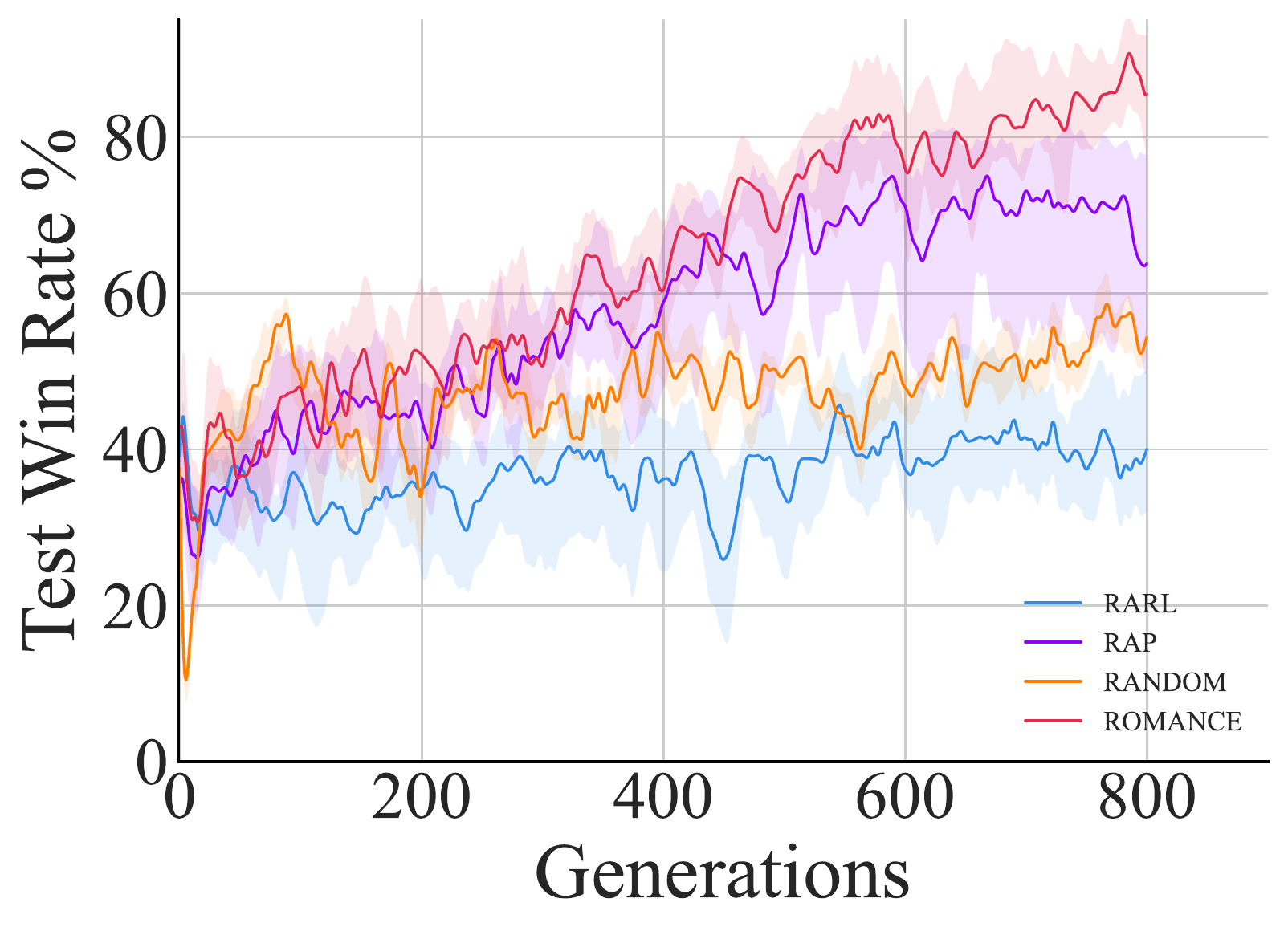}
		\centerline{(c) MMM}
	    \end{minipage}
	    \begin{minipage}{0.492\linewidth}
		\centering
		\includegraphics[width=1\linewidth]{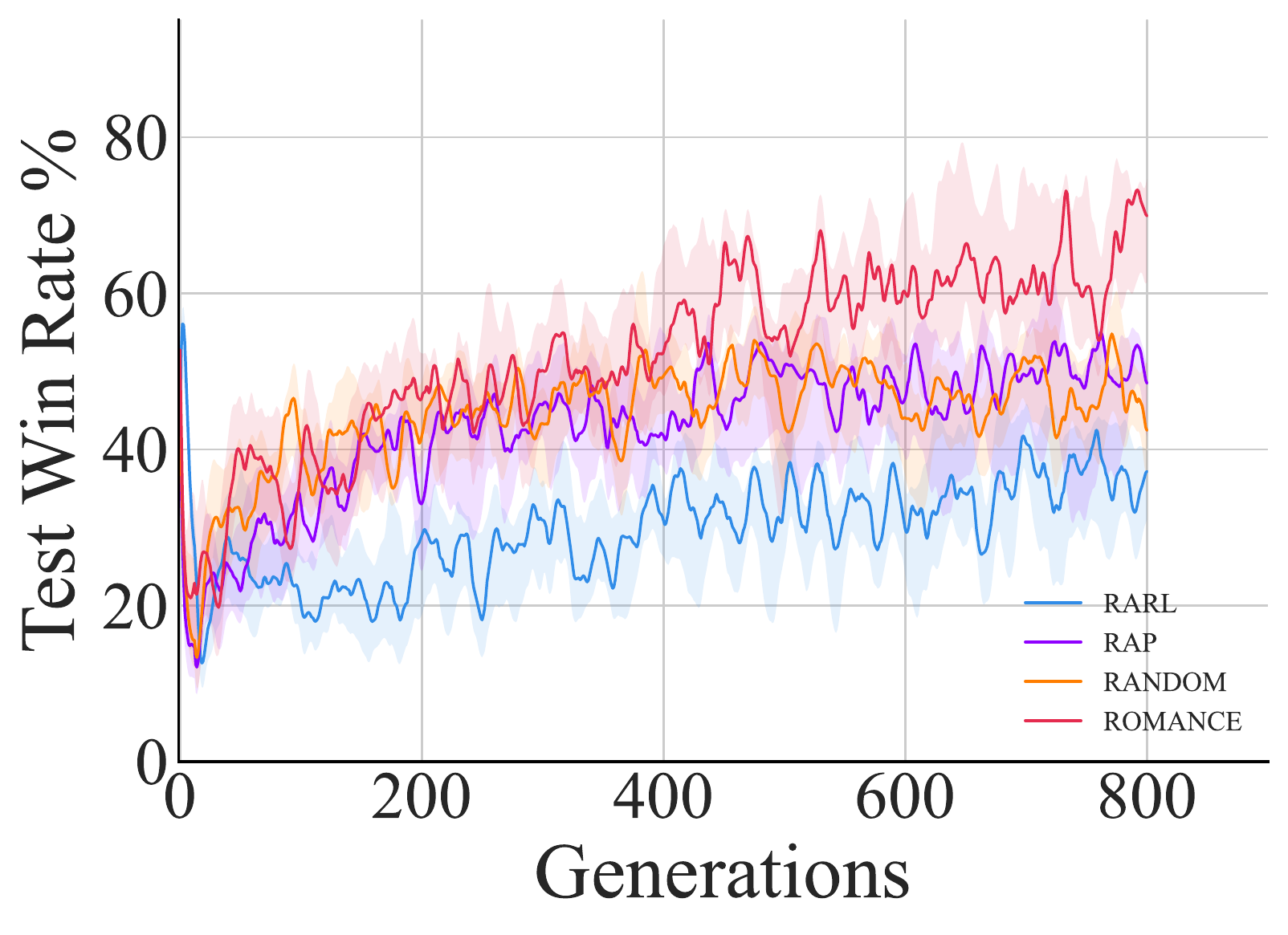}
		\centerline{(d) 1c3s5z}
	    \end{minipage}
    \caption{Average test win rates on four more maps. }
    \label{curve}
\end{figure}

Besides what has been presented in the main  manuscript, we show the learning curve of different methods (built on QMIX) on other four maps in Fig.~\ref{curve}. We can observe that ROMANCE achieves the best performance under the strong adversarial attack on other maps. We can also observe that RARL even performs worse than RANDOM in map MMM and 1c3s5z. We believe that when ther exist many heterogeneous agents, the ego-system might be overfitting to a specific type of attackers easily. 
In Fig.~\ref{vdn_qplex_curve}, we present the learning curves of different methods implemented on QPLEX and VDN on map 2s3z. The curves show that ROMANCE can
significantly enhance the robustness of value-based MARL algorithms when they are integrated.

\begin{figure}[h!]
    \centering
	    \begin{minipage}{0.492\linewidth}
		\centering
		\includegraphics[width=1\linewidth]{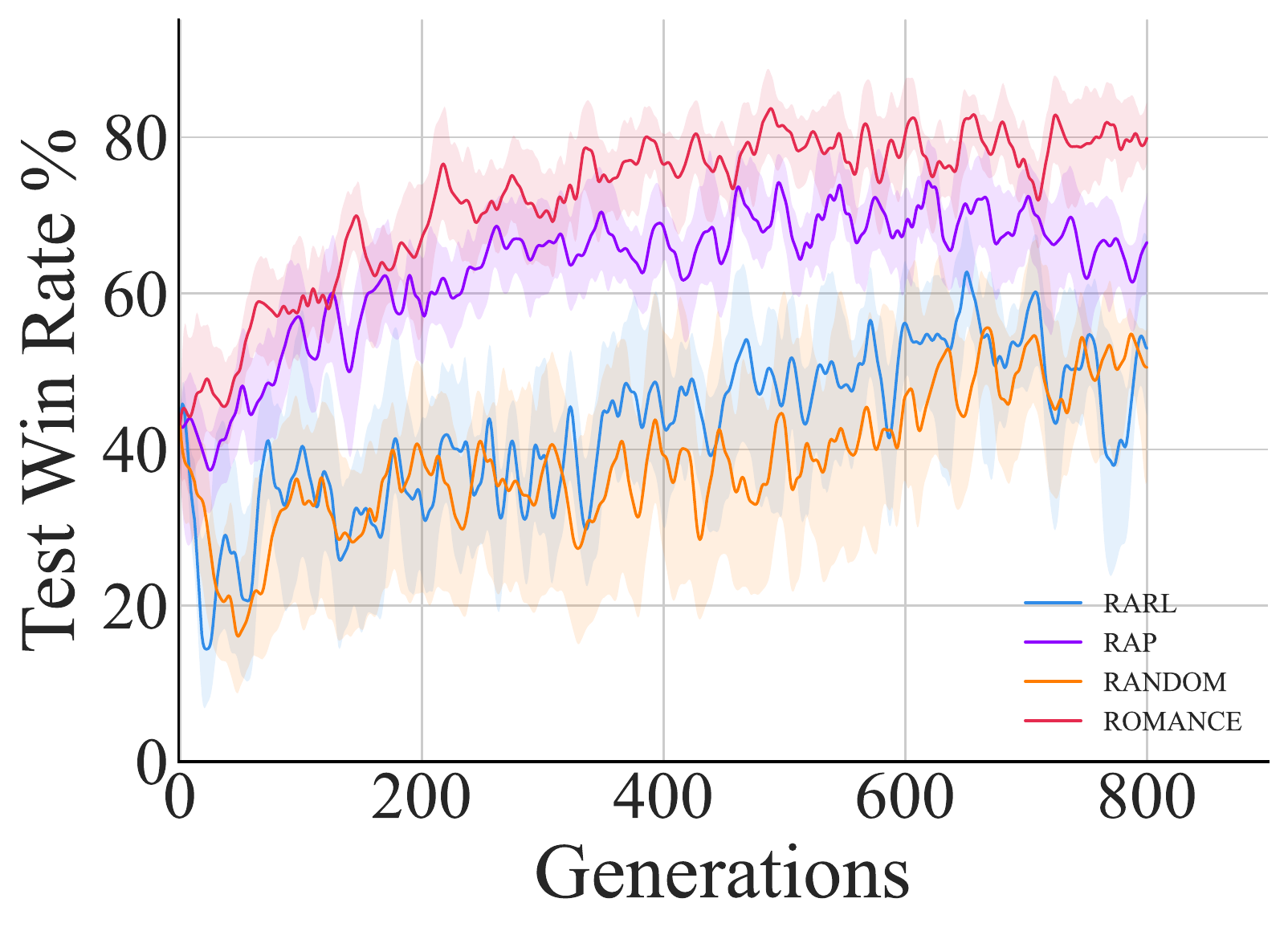}
		\centerline{(a) VDN}
	    \end{minipage}
	    \begin{minipage}{0.492\linewidth}
		\centering
		\includegraphics[width=1\linewidth]{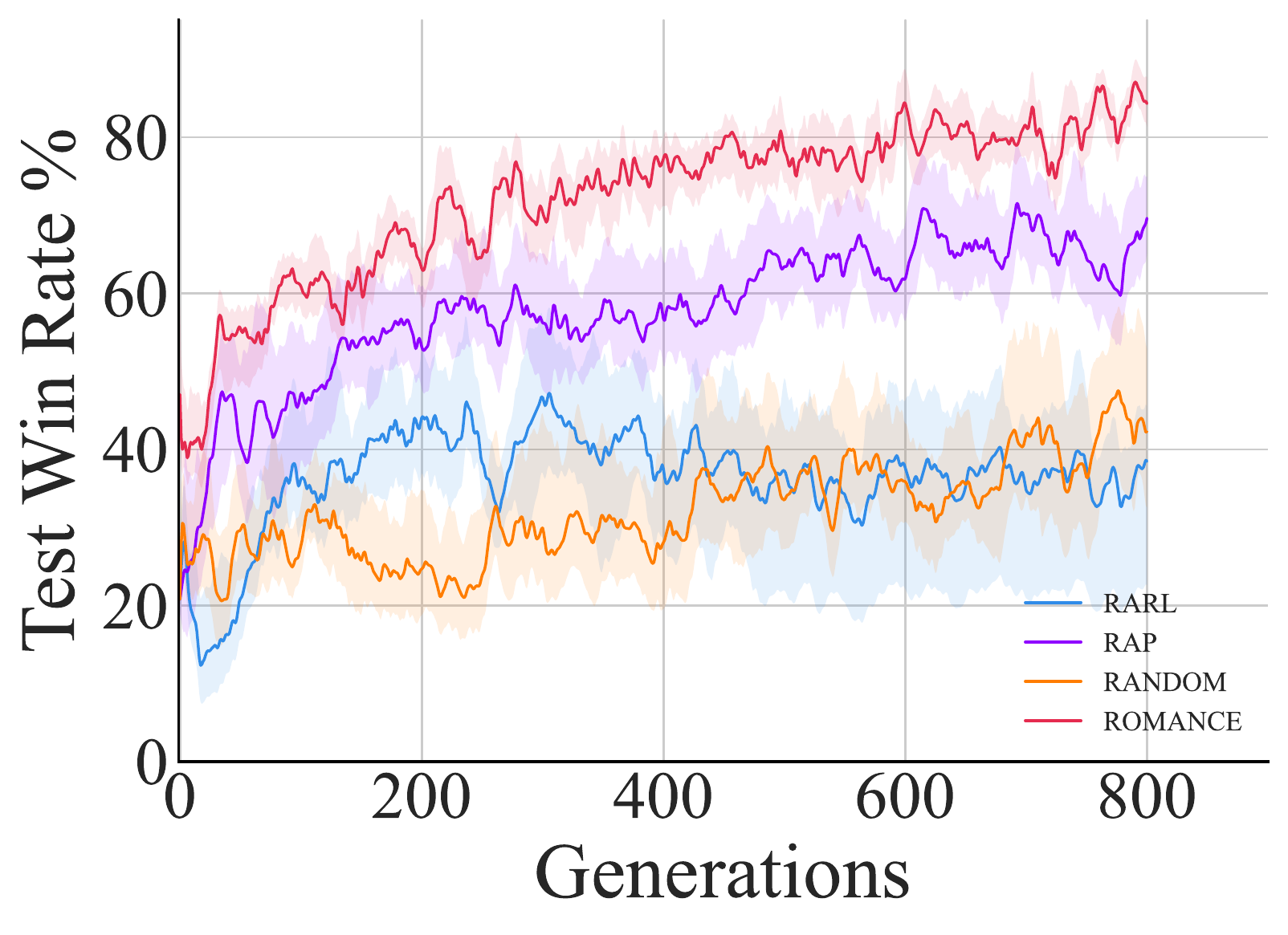}
		\centerline{(b) QPLEX}
	    \end{minipage}
	\caption{Average test win rates of VDN and QPLEX on map 2s3z, the result of QMIX is shown in the main manuscript.}
    \label{vdn_qplex_curve}
\end{figure}

\begin{figure}[t!]
    \centering
    	\begin{minipage}{0.49\linewidth}
		\centering
		\includegraphics[width=1\linewidth]{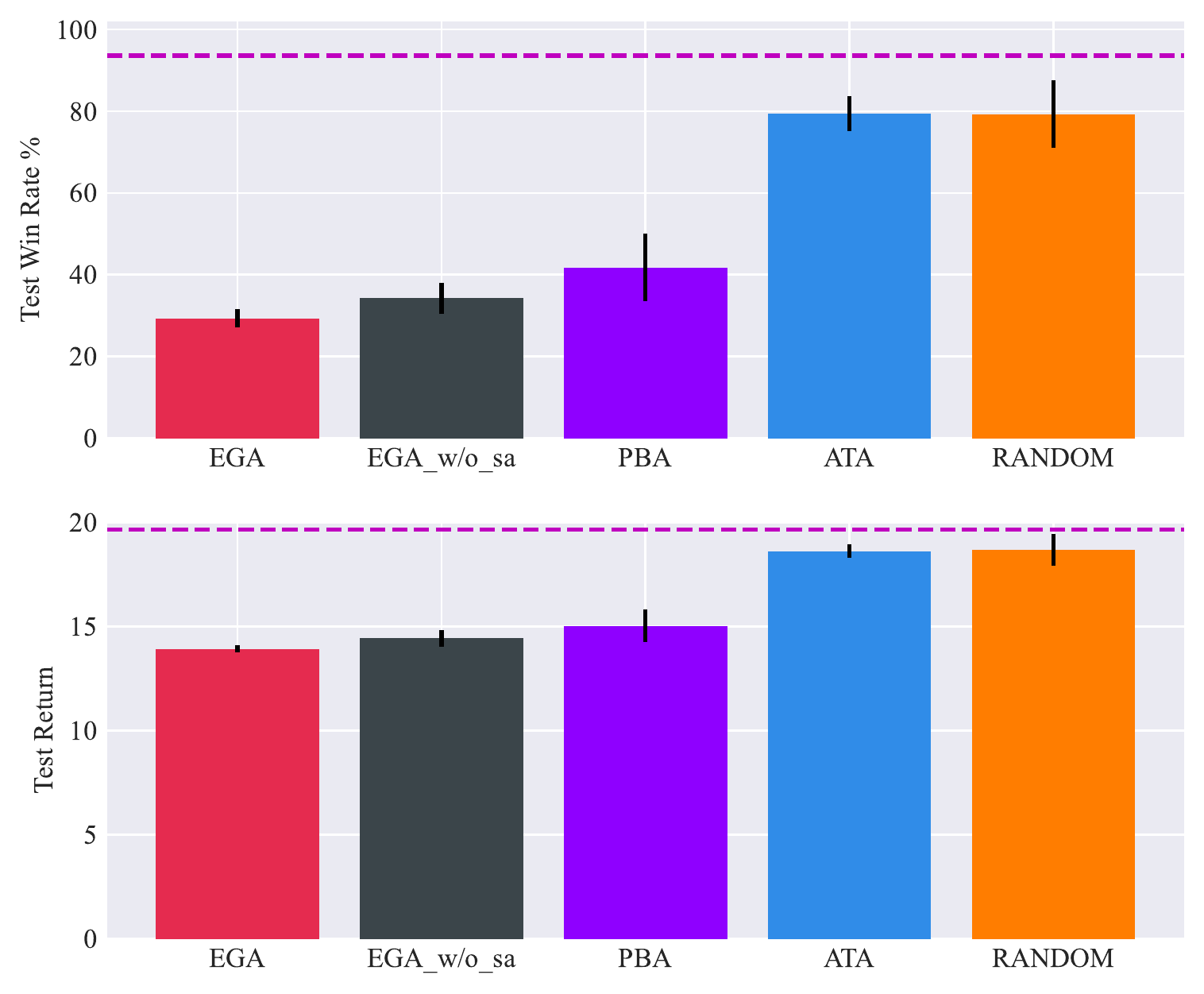}
		\centerline{(a) 2s3z}
	    \end{minipage}
	    \begin{minipage}{0.49\linewidth}
		\centering
		\includegraphics[width=1\linewidth]{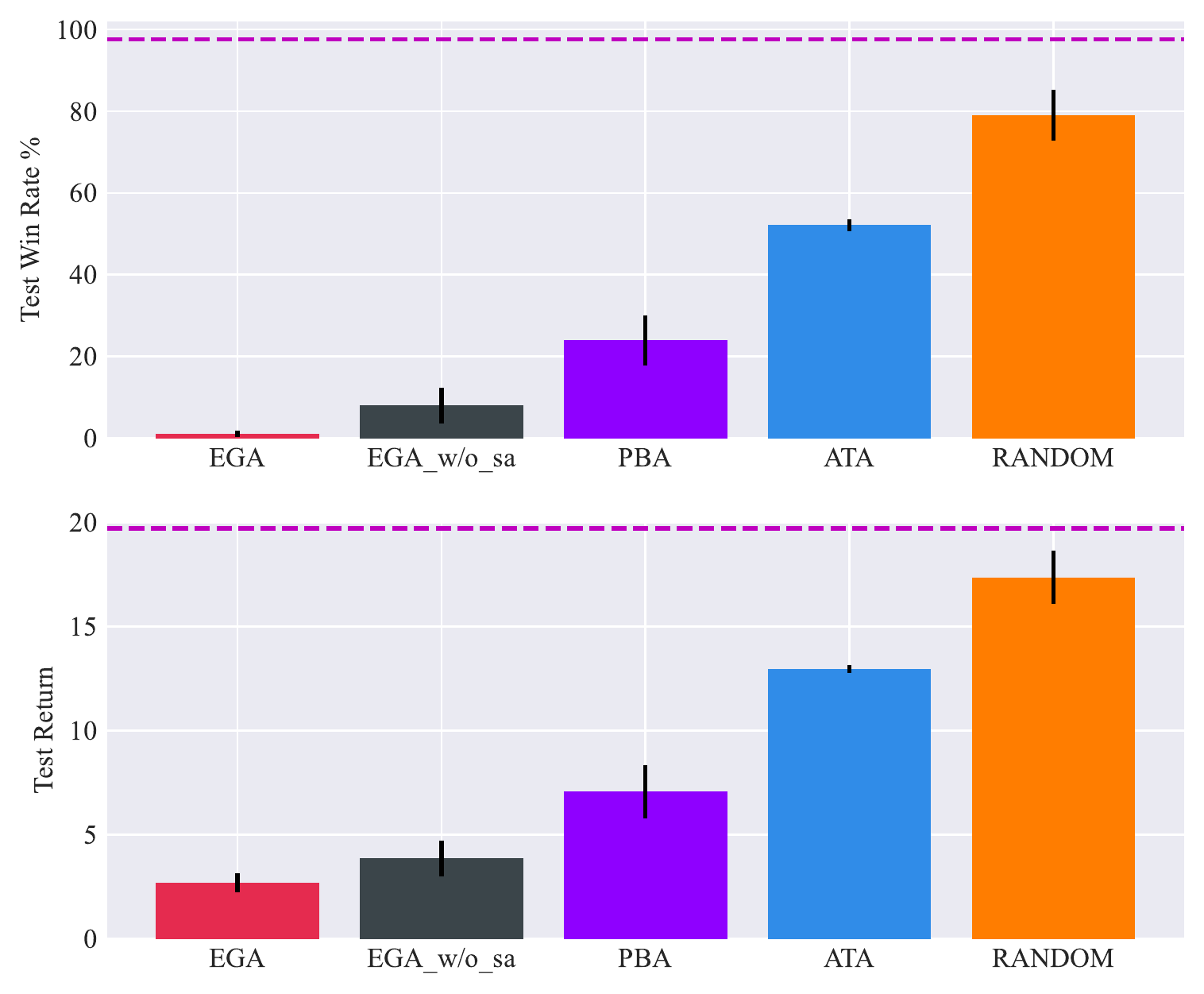}
		\centerline{(b) 3m}
	    \end{minipage}
	    
	    \begin{minipage}{0.49\linewidth}
		\centering
		\includegraphics[width=1\linewidth]{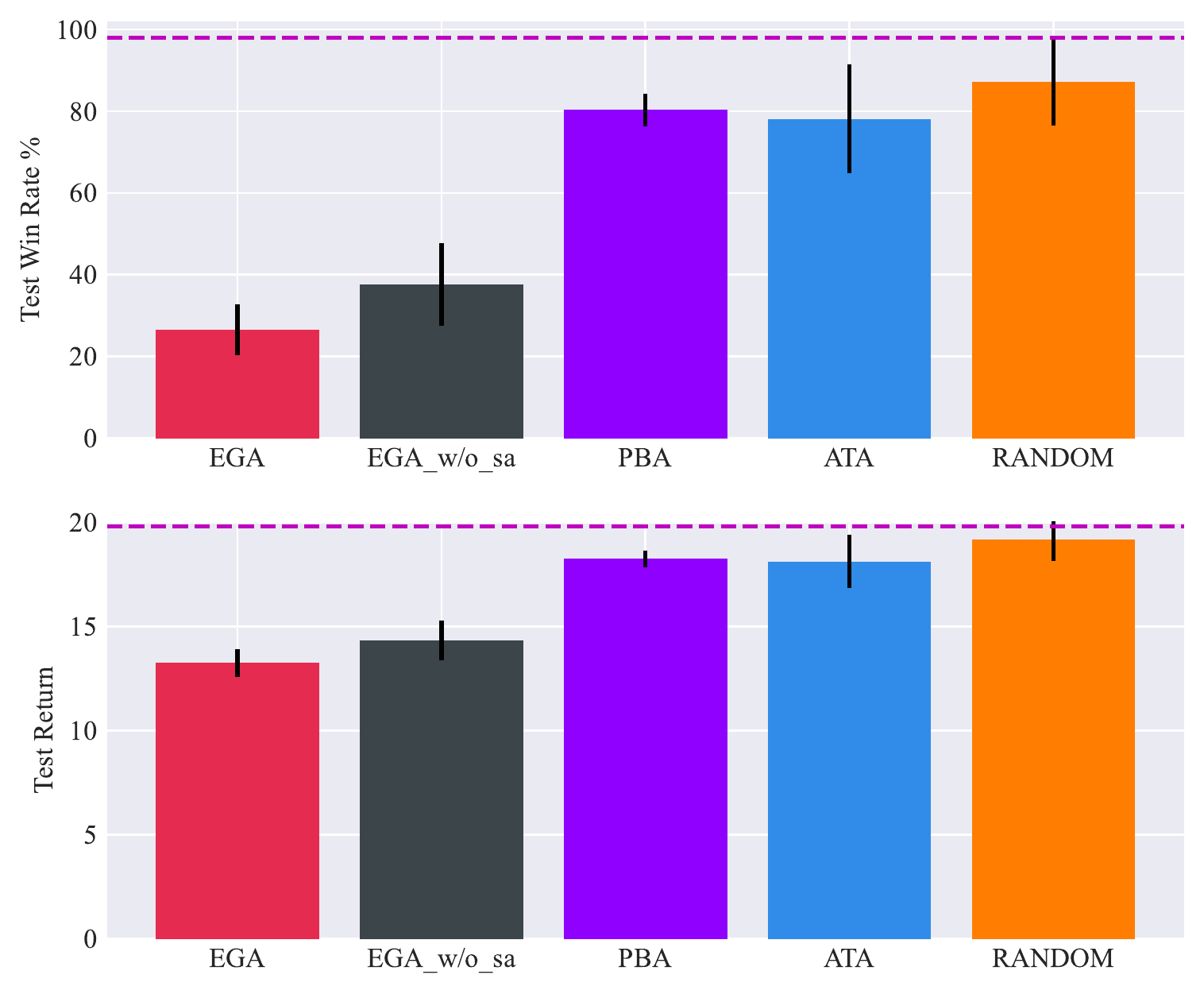}
		\centerline{(c) 3s\_vs\_3z}
	    \end{minipage}
	    \begin{minipage}{0.49\linewidth}
		\centering
		\includegraphics[width=1\linewidth]{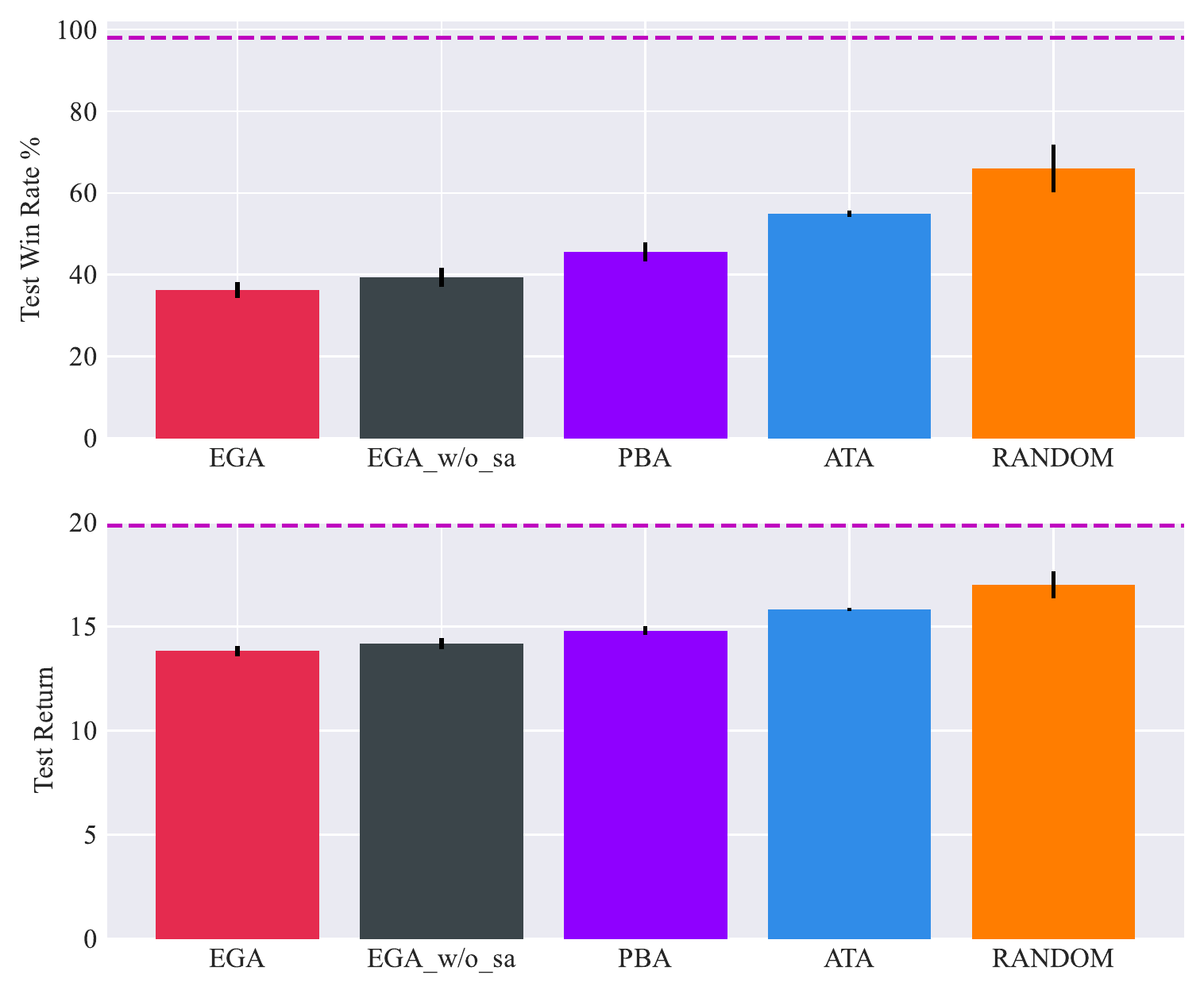}
		\centerline{(d) 8m}
	    \end{minipage}
	    
	    \begin{minipage}{0.49\linewidth}
		\centering
		\includegraphics[width=1\linewidth]{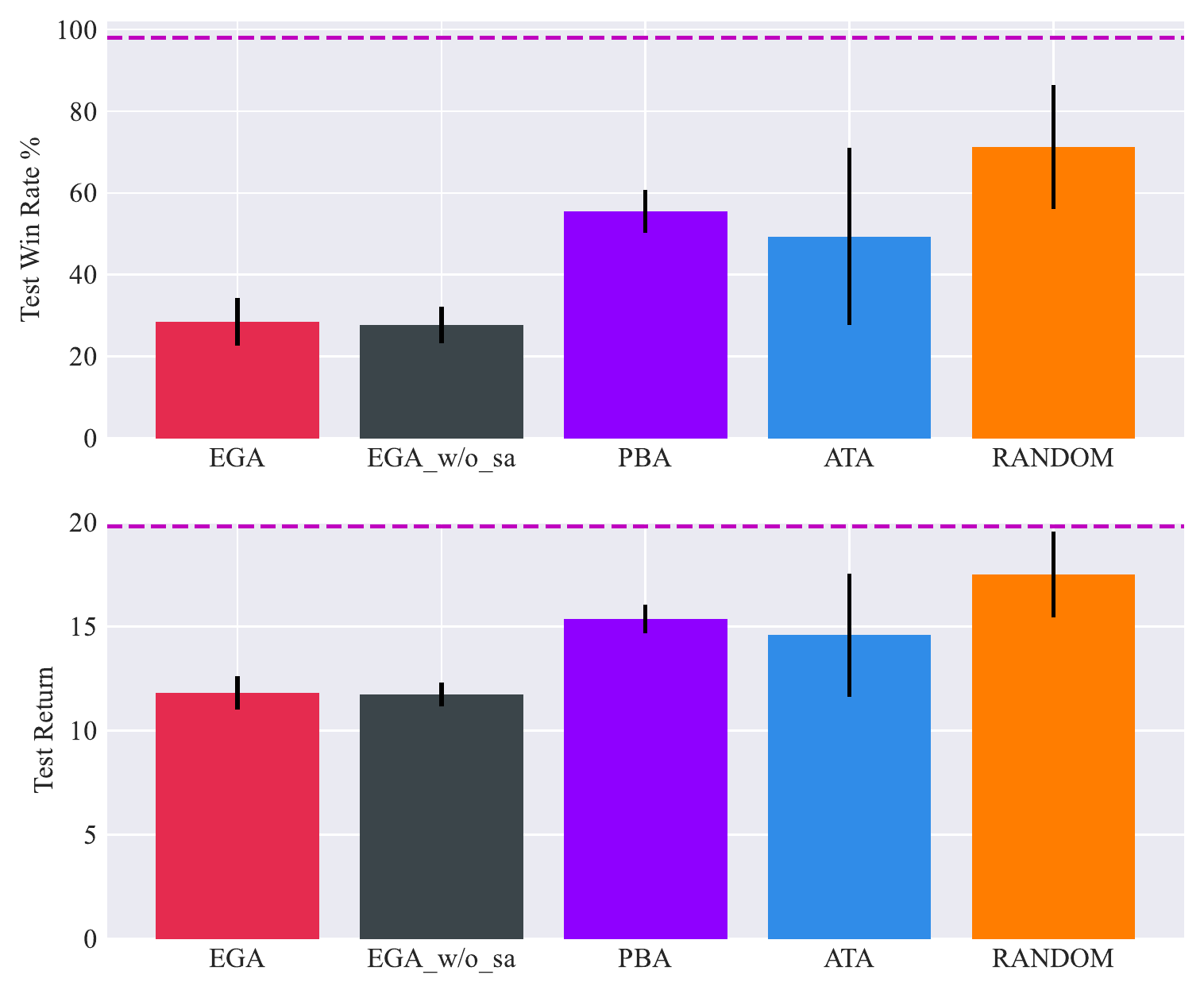}
		\centerline{(e) MMM}
	    \end{minipage}
	    \begin{minipage}{0.49\linewidth}
		\centering
		\includegraphics[width=1\linewidth]{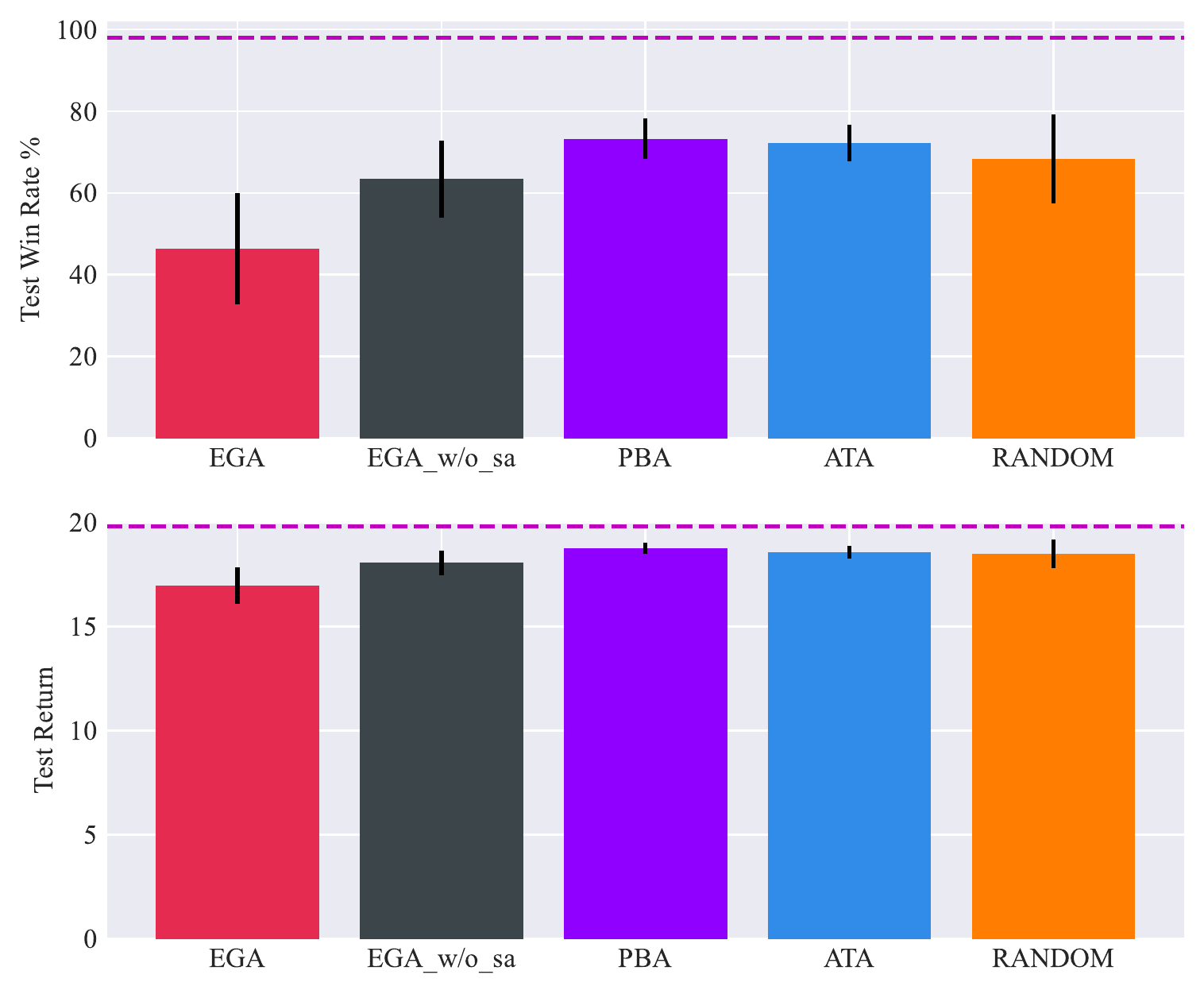}
		\centerline{(f) 1c3s5z}
	    \end{minipage}
	    \caption{The quality of attackers generated by different methods on more six maps.}
	    \label{barattack}
\end{figure}

In Fig.~\ref{barattack} we show the attackers' quality generated by different methods on all six maps. For EGA, EGA\_w/o\_sa and PBA that generate more than one attackers in a single run, we take the mean value of five best attackers as the quality of each run. The superiority of EGA over other methods in all six maps demonstrates its effectiveness.

In Tab.~\ref{tab3}-\ref{tab5}, we investigate how different types of hyperparameters affect the performance of ROMANCE. Respectively, we alter the size of the archive, size of the population, and the number of attack during training and then test the ego-system under three settings. 

\textbf{Archive Size.}\quad
The archive is used to record the attackers with high-performance and diverse behaviors generated so far. It is crucial for ROMANCE, as new attackers are generated based on those selected from the archive. Too small size results in decrease in diversity, but too large size makes it inefficient to select high-performing attackers. 

As shown in Tab.~\ref{tab3}, we find that the slightly bigger archive size promotes the improvement of robustness as its larger capacity allows for maintaining more high-performing and diverse attackers. The performance might also decrease if it goes too large, with a decrease in efficiency, as explained above.
\begin{table}[h!]
\centering
\resizebox{0.474\textwidth}{!}{
\begin{tabular}{c|c|c|c} 
\hline
Archive Size                        & Natural        & Random Attack  & EGA             \\ 
\hline
11                                   & $94.0\pm9.06$ & $83.7\pm10.3$ & $71.2\pm12.1$  \\
13                                   & $98.5\pm0.68$ & $88.9\pm1.59$ & $76.5\pm4.16$  \\
\rowcolor[rgb]{0.893,0.893,0.893} 15 & $97.9\pm1.34$ & $89.1\pm1.97$ & $81.6\pm0.84$  \\
17                                   & $98.1\pm0.62$ & $88.2\pm0.94$ & $83.6\pm3.02$  \\
19                                   & $96.8\pm1.14$ & $86.6\pm1.06$ & $78.8\pm7.40$  \\
\hline
\end{tabular}}
\caption{Average test win rate of ROMANCE on map 2s3z when archive size changes.}
\label{tab3}
\end{table}

\textbf{Population Size.}\quad Tab.~\ref{tab4} describes the robustness of ROMANCE when the size of the population changes. The population size refers to the number of attackers the ego-system will encounter in one generation. A larger population size promotes the robustness of the ego-system in the final stage, but might also generate attackers with similar behavior, thus harming the efficiency. Despite the same archive size, training the best response to a small size population with only 2 or 3 attackers tends to result in overfitting, and thus weakens the generalization ability against diverse attackers. Results in Tab.~\ref{tab4} show that the default population size $4$ is appropriate in the map 2s3z.

\begin{table}[h!]
\centering
\resizebox{0.474\textwidth}{!}{
\begin{tabular}{c|c|c|c} 
\hline
Population Size                           & Natural         & Random Attack  & EGA             \\ 
\hline
2                                   & $94.7\pm4.56$  & $83.4\pm7.40$ & $66.2\pm9.44$  \\
3                                   & $92.6\pm5.47$ & $82.0\pm9.78$ & $68.8\pm12.9$  \\
\rowcolor[rgb]{0.893,0.893,0.893} 4 & $97.9\pm1.34$ & $89.1\pm1.97$ & $81.6\pm0.84$   \\
5                                   & $98.4\pm0.83$  & $88.7\pm2.74$  & $79.2\pm5.25$   \\
6                                   & $98.8\pm0.45$  & $89.6\pm1.07$  & $80.9\pm1.34$   \\
\hline
\end{tabular}}
\caption{Average test win rate of ROMANCE on map 2s3z when population size changes.}
\label{tab4}
\end{table}

\textbf{Training Attack Budget.}\quad We also present ROMANCE's average test win rate under different budgetary training attack numbers in Tab.~\ref{tab5}. When the training budget is less than the testing one, the generalization ability of our method guarantees their decent performance. However, it is still inferior to the ego-system whose training budgetary attack number is in accordance with the testing one. Excess training attack numbers might also cause performance degradation, as too strong attackers bring a conservative ego-system.

\begin{table}[htbp!]
\centering
\resizebox{0.474\textwidth}{!}{
\begin{tabular}{c|c|c|c} 
\hline
Attack Num                                   & Natural        & Random Attack  & EGA             \\ 
\hline
6                                   & $97.3\pm1.42$ & $87.2\pm2.42$ & $71.6\pm5.59$  \\
7                                   & $98.0\pm1.05$ & $87.2\pm3.28$ & $76.2\pm2.80$  \\
\rowcolor[rgb]{0.893,0.893,0.893} 8 & $97.9\pm1.34$ & $89.1\pm1.97$ & $81.6\pm0.84$  \\
9                                   & $97.8\pm1.38$ & $88.2\pm3.29$ & $84.4\pm2.68$  \\
10                                  & $98.3\pm0.86$ & $87.4\pm2.91$ & $80.1\pm1.99$  \\
\hline
\end{tabular}}
\caption{Average test win rate of ROMANCE on map 2s3z when attacker number changes.}
\label{tab5}
\end{table}

\bibliography{romance}
\bibliographystyle{plainnat}
\end{document}